\def\lae{\mathrel{<\kern-1.0em\lower0.9ex\hbox{$\sim$}}}
\def\gae{\mathrel{>\kern-1.0em\lower0.9ex\hbox{$\sim$}}}

\documentclass{emulateapj}

\usepackage{graphicx}

\slugcomment{Accepted for publication in the Astronomical Journal}

\lefthead{{\sc Jord\'an et al.}}
\righthead{Globular Clusters Systems of cD Galaxies}

\begin{document}

\title{{\it Hubble Space Telescope} Observations of cD Galaxies and their Globular
Cluster Systems\altaffilmark{1,2}}

\author{Andr\'es Jord\'an\altaffilmark{3}, Patrick C\^ot\'e}
\affil{Department of Physics and Astronomy, Rutgers University, Piscataway, NJ 08854 \\ 
andresj@physics.rutgers.edu, pcote@physics.rutgers.edu}
\medskip

\author{Michael J. West}
\affil{Department of Physics \& Astronomy, University of Hawaii, Hilo, HI 96720 \\ 
west@bohr.uhh.hawaii.edu}
\medskip

\author{Ronald O. Marzke}

\affil{Department of Physics \& Astronomy, San Francisco State University, 
1600 Holloway Avenue, San Francisco, CA 94132 \\ marzke@quark.sfsu.edu}
\medskip

\author{Dante Minniti}
\affil{Departamento de Astronom\'{\i}a y Astrof\'{\i}sica,
P. Universidad Cat\'olica, Casilla 104, Santiago 22, Chile \\ dante@astro.puc.cl}

\and

\author{Marina Rejkuba}
\affil{European Southern Observatory, Karl-Schwarzschild-Str. 2, 85748 Garching, Germany \\
mrejkuba@eso.org}

\medskip

\altaffiltext{1}{Based on observations with the NASA/ESA {\sl Hubble Space Telescope}
obtained at the Space Telescope Science Institute, which is operated by the Association
of Universities for Research in Astronomy, Inc., under NASA contract NAS 5-26555}
\altaffiltext{2}{Based in part on observations obtained at the European Southern Observatory, 
for the VLT program 68.D-0130(A).
}
\altaffiltext{3}{Claudio Anguita Fellow}

\begin{abstract}
We have used WFPC2 on the {\sl Hubble Space Telescope} to obtain F450W and 
F814W images of four cD galaxies (NGC~541 in Abell~194, NGC~2832 in Abell~779, 
NGC~4839 in Abell~1656 and NGC~7768 in Abell~2666) 
in the range $5400 \lesssim cz \lesssim$ $8100$ km s$^{-1}$.  For NGC~541, 
the {\sl HST} data are supplemented by ground-based $B$ and $I$ images 
obtained with the FORS1 on the VLT. We present surface brightness and
color profiles for each of the four galaxies, confirming their classification
as cD galaxies. Isophotal analyses 
reveal the presence of subarcsecond-scale dust 
disks in the nuclei of NGC~541 and NGC~7768. 
Despite the extreme nature of these galaxies in terms of spatial extent
and luminosity, our analysis of their globular cluster systems 
reveals no anomalies in terms of specific frequencies,
metallicity gradients, average metallicities, or the metallicity offset
between the globulars and the host galaxy. We show that the latter
offset appears roughly constant at $\Delta\mbox{[Fe/H]}\sim$ 0.8~dex for 
early-type galaxies spanning a luminosity range of roughly four orders of magnitude.
We combine the globular cluster metallicity 
distributions with an empirical technique described in a series of earlier
papers to investigate the form of the protogalactic mass spectrum 
in these cD galaxies. We find that the observed GC metallicity
distributions are consistent with those expected if cD
galaxies form through the cannibalism of numerous galaxies
and protogalactic fragments which formed their stars and
globular clusters before capture and disruption. However,
the properties of their GC systems suggest that dynamical
friction is {\it not} the primary mechanism by which these galaxies are
assembled. We argue that cDs instead form rapidly,
via hierarchical merging, prior to cluster virialization.
\end{abstract}


\section{Introduction}
\label{sec:intro}

In spatial extent and luminosity, cD galaxies are intermediate between normal 
galaxies and galaxy clusters. Indeed, the cD classification itself was introduced 
in the work of Matthews, Morgan \& Schmidt (1964) to denote the very large
D galaxies that they found in some clusters. (The `c' prefix was borrowed from 
the notation for supergiant stars in stellar spectroscopy.) Although the
extraordinary sizes of cD galaxies was readily apparent to Matthews et~al. (1964), 
the outstanding nature of their distinctive envelopes was fully appreciated
only after the deep photographic surface photometry of Oemler (1976) and 
Schombert (1986; 1988).

In this paper, we adopt the definition of a cD as given in Schombert (1986) and Tonry 
(1987). That is, cDs are elliptical galaxies with shallow surface brightness profiles
$$d\log\mu_V/d\log r \approx -2 \eqno{(1)}$$ 
at $\mu_V \approx 24$ mag arcsec$^{-2}$. These galaxies exhibit a characteristic `break' 
over an $r^{1/4}$ law, beginning at $r_b \approx 10-100$~kpc, and are
much brighter than typical elliptical galaxies, with luminosities $\sim$ 
10$L^*$ (Sandage \& Hardy 1973; Schombert 1986).
cD galaxies are always found in dense regions,
and in virtually all cases, they are 
located near the spatial and kinematical center of their host cluster, or subcluster.
These facts would seem to suggest that the formation of cD galaxies is unique to 
the cluster environment and is closely linked to its dynamical history.

There are, in fact, reasons to believe that the envelopes themselves are
distinct entities from the galaxies themselves. First, cD envelope luminosity 
is weakly correlated with some properties of the host cluster, most notably with
cluster richness and  X-ray luminosity (Schombert 1988). Second, the envelopes 
have surface brightness profiles with power-law slopes that are similar to those
measured from the surface density profiles of the surrounding cluster galaxies.
Finally, both the position angle and ellipticity of cD galaxy isophotes commonly
show discontinuities at $r_b$, where the envelope begins to dominate the surface 
brightness profile (Schombert 1988; Porter, Schneider \& Hoessel 1991).

Models for the formation of cD galaxies have tended to focus on dynamical effects that, 
in dense environments, might produce extended low-surface-brightness envelopes.
A tidal stripping origin was first proposed by Gallagher \& Ostriker (1972), and 
examined in detail by Richstone (1976). In this scenario, close encounters
between galaxies liberate material which surrounds the central galaxy;
if the velocity of the encounter is sufficiently low, the interaction may
result in a merger. The most massive galaxies would be preferentially depleted as they are most
strongly affected by dynamical friction. If many such mergers occur, then the 
central galaxy may grow primarily through this process of ``cannibalism".  The rates 
for galactic cannibalism in rich clusters were computed first with approximate 
theories that used average interaction rates without considering the actual 
galaxy trajectories (Ostriker \& Tremaine 1975; White 1976), and later with 
Monte Carlo techniques (Hausman \& Ostriker 1978). These calculations predicted 
rather high cannibalism rates in rich clusters, leading to a luminosity 
growth of the first ranked of $\gae10L^{*}$ over the lifetime of the host cluster. 
The finding that $\sim$ 50\% of brightest cluster galaxies (BCGs) contain multiple 
nuclei ($e.g.$, Hoessel \& Schneider 1985) has sometimes been interpreted as 
supporting evidence for this model. 

Richstone \& Malumuth (1983) carried out more detailed simulations of the
growth of central galaxies by following the trajectories of individual
galaxies in a fixed cluster potential. Approximately 30\% 
of their simulations produced very massive galaxies at the cluster center, which 
they identified as cD galaxies. In their simulations, which treated dynamical
friction, mergers and tidal stripping simultaneously, the growth of the cD
occurred throughout the lifetime of the cluster. Merritt (1984; 1985) reached
a rather different conclusion using a statistical description of the 
galaxy orbital energy and mass distribution, and a Fokker-Planck treatment 
of the evolution of these distributions due to tidal stripping and dynamical 
friction. He found that neither tidal stripping nor mergers occur at a 
significant rate in rich clusters after virialization, and concluded that 
the morphology of BCGs in a rich cluster --- and of cDs in particular --- 
is fixed during the early stages of cluster evolution, probably during 
the collapse and virialization of compact groups or poor clusters.
In the post-collapse stages of evolution, the central galaxy was found
to grow in luminosity by only $\sim 1L^*$. Dynamical analysis of galaxies 
with multiple-nuclei led Merrifield \& 
Kent (1991) and Blakeslee \& Tonry (1992) to conclude such galaxies 
have probably increased their luminosities by no more than $\approx$ 1-2$L^{*}$ over a
typical cluster lifetime as a result of galactic cannibalism. These findings 
are consistent with the modest post-virialization growth rates advocated by 
Merritt (1985).

In principle, the formation of central galaxies in cluster and group
environments is best studied with high-resolution, N-body simulations
($e.g.$, Funato, Makino \& Ebisuzaki 1993; Bode et~al. 1994; Garijo, 
Athanassoula \& Garc\'{\i}a-G\'omez 1997). The results of such simulations,
however, depend sensitively on the assumed initial conditions,
so it is important to start with a physically-motivated paradigm of 
structure formation. The first attempt to do so was by Dubinski (1998), who
followed the formation and evolution of the central BCG within a single
rich cluster chosen from a cold-dark-matter simulation. In his simulation,
which included both stellar and dark matter components, a massive central
central galaxy formed via the merger of several massive ``protogalaxies" 
early in the cluster history. Due to the filamentary structure of the cluster
at these times, the mergers themselves were found to be highly anisotropic, with 
the protogalaxies infalling radially along with filaments. In agreement with
previous findings, Dubinski (1998) found
that cannibalism by dynamical friction during the post-collapse evolution
accounts for only a small fraction of the accreted mass. It is worth
noting, however, that the central galaxy in his simulation did not
develop the extended envelope that is characteristic of cD galaxies.

Another, quite different, formation model for cD galaxies involves cooling 
flows (see, $e.g.$, Fabian 1994). Since the radiative cooling times for 
intracluster gas are short enough that gas can cool and settle to the cluster 
center, it has been suggested that cD envelopes may arise from the gradual 
deposition of this cool gas.  However, this possibility now seems 
remote in light of {\sl XMM-Newton} observations that the X-ray gas does 
not cool significantly below a threshold temperature of $kT \sim$ 1-2 keV 
(Kaastra et~al. 2001; Peterson et~al. 2001; Tamura et~al. 2001). 

Clearly, an understanding of the mass distribution of the galaxies and protogalaxies
from which cD galaxies were assembled would be helpful in discriminating between 
the various models.  Globular clusters (GCs) are useful in this regard since 
they are bright (with a mean absolute magnitude of $M_V \simeq -7.3$ mag; 
Harris 2001) and are found in large numbers around massive, early-type galaxies. 
Moreover, observations of nearby dwarfs and giant ellipticals have shown that there is a 
tight correlation between mean GC metallicity and the total galaxy luminosity 
($e.g.$, Brodie \& Huchra 1991). Thus, GCs should be powerful diagnostics of 
the cD formation process since the total luminosities and masses of the 
cannabilized galaxies (assuming cD galaxies formed in this way) 
must be imprinted in the metallicities of their present-day GC systems.
Yet color (and thus metallicity) information is available in the literaure for 
the GC systems of only five cD
galaxies\footnote{Based on the compilation of van den Bergh (2001). Note that we
include M87 as a cD galaxy (de Vaucouleurs \& Nieto 1978) and have added 
NGC~6166 (Bridges et~al. 1996) to his list.}.

In this paper, we present $B$ and $I$ photometry of the GC systems of four distant cD 
galaxies and place 
constraints on the distribution of progenitor masses. This sample nearly doubles
the number of cD galaxies with GC metallicity distributions.
The paper is organized as follows. In \S 2, we describe the observations and data 
reduction procedures. \S 3 presents a summary of the observed properties of the 
galaxies and their GC systems. In \S 4, we present an analysis of the observed GC
metallicity distribution functions. We summarize our findings and draw conclusions
in \S 5.

\section{Observations and Data Reduction}

We used WFPC2 on board {\sl HST} to obtain F450W (broadband $B$)
and F814W (broadband $I$) images of four cD galaxies: NGC~541, NGC~2832, 
NGC~4839 and NGC~7768.  The observations, corresponding to program GO-8184, 
are summarized in Table~\ref{tab:log}. Three of the galaxies (NGC~2832, 
NGC~4839 and NGC~7768) were selected from the sample of 
Schombert (1986). The fourth galaxy, NGC~541, was selected from the
compilation of Bird (1994), although it has in the past been variously 
classified as D or cD ($e.g.$, Dressler 1980; Bird 1994). To clarify 
the issue of its morphological type, we subsequently obtained $B$ and $I$ 
images of NGC~541 with the VLT. These data and our analysis of them are 
presented in \S \ref{sec:vlthst}. Properties of the host clusters for 
each of our program galaxies are summarized in 
Table~\ref{tab:prop_clus}.\footnote{Throughout this paper, we identify 
NGC~4938 with the subcluster projected $\approx$ 1.1~Mpc to the southwest 
of the Coma cluster center.}

The raw WFPC2 data were processed with the standard STScI pipeline using the
best available calibration files. We verified  that
the images were properly aligned with cross-correlation measurements of the
shifts using the DITHER package in IRAF\footnote{IRAF is distributed by the National 
Optical Astronomy Observatories, which are operated by the Association of Universities 
for Research in Astronomy, Inc., under cooperative agreement with the National 
Science Foundation.}. Cosmic rays were rejected by combining images of 
a given galaxy and filter using the IRAF task GCOMBINE.

\subsection{Surface Photometry}
\label{sec:sb}

The images were ``mosaiced'' using the task WMOSAIC in the STSDAS package. Bad
pixels, regions affected by vignetting, and any remaining cosmic rays were masked. 
The images were visually inspected and any bright contaminating objects
present in the fields were also masked.  Surface photometry was performed 
with the task ELLIPSE, which is based on an algorithm by Jedrzejewski (1987). 
The algorithm was run on the $I$ image and the resulting solution  was 
transferred to the $B$ band image, where ELLIPSE was run in ``no-fit'' mode. 
The conversion to Johnson $B$ and $I$ magntiudes was done in the same manner
as for the point sources  (see \S\ref{sec:point}) except that an additional
$0.1$ mag  correction was applied to correct the 
zeropoints to infinite aperture (Holtzman et~al. 1995).

A common difficulty in performing surface photometry with WFPC2 is the
limited field of view, which greatly complicates the estimation of the 
background sky brightness. Simply stated, the galaxies are still apparent
even in the outskirts of our images.  One approach to this problem is to 
fit a parametric curve (such as a de Vaucouleurs
law or a power-law model) plus some constant background to a section of the 
profile, and estimate the sky directly from the fit. In our case,
such a procedure is unsatisfactory. For the cD galaxies examined here,
it is risky to assume a specific functional form when there is no
{\sl a priori} knowledge of where envelope will appear.  Furthermore,
sky measurements arrived at in this way will have independent errors in
different bandpasses, potentially biasing the inferred color profiles at 
faint levels. Finally, as emphasized by Schombert (1986), no parametric 
function can capture the peculiarities of the profiles that may 
affect the sky estimates. 

In an attempt to alleviate these problems, we adopted the following procedure. 
In the outskirts of our fields, the galaxy contribution to the total intensity, 
though non-negligible, is nevertheless small. By estimating the ratio of the 
sky intensities directly from the {\sl ratio} image of the $B$ and $I$ images, 
it is possible to reduce the uncertainty in the estimated sky color.
Let us denote the counts due to the sky in the $I$ and $B$ bandpasses by 
$S_I$ and $S_B$, respectively, and the corresponding intensities from the 
galaxy light by $G_I$ and $G_B$. Quite generally, the galaxy intensities in
the different bandpasses will be certain 
fractions of the sky intensities: $i.e.$, $G_I=\alpha S_I$ and $G_B=\beta S_B$. 
For a typical giant elliptical, $(B-I)_{gal}\sim2.2$ ($e.g.$, Fukugita, 
Shimasaku \& Ichikawa 1995; see below also) whereas the color of the ``sky" seen 
by {\sl HST} is $(B-I)_{sky}\sim1.5$ (\textit{HST} Data Handbook). 
In general, we then have
$$\log(q) \equiv \log(\beta/\alpha) = 0.4\biggl [(B-I)_{sky}-(B-I)_{gal} \biggr ] \eqno{(2)}$$
and, if we estimate the ratio, $R$, of the sky intensities in the outskirts of the fields, 
$$R=\frac{S_I+\alpha S_I}{S_B+\beta S_B} = \frac{S_I}{S_B}\left(\frac{1+\alpha}{1+q\alpha}\right)\eqno{(3)}$$
The value of $R$ was calculated for a range in plausible values of $\alpha$ 
and $q$. The distribution of $R$ is sharply peaked at $R\sim 1.023 S_I/S_B$, 
with an interquartile range of $\approx 0.01$ and a median of $R=1.025S_I/S_B$.
We therefore take $R/1.025$ for our estimate of $S_I/S_B$, with $R$ 
estimated directly from the ratio image. 

Since the galaxies are fainter with respect to the sky in the bluer bandpass, 
$S_B$ was determined from the outer regions of the $B$ images, along the direction 
where the profile declines most steeply. The sky brightness in $I$ was
then obtained from $S_I=S_BR/1.025$. Note that in the regions where $S_B$ was measured,
the galaxy contribution is at least a factor of 20 less than that of the 
sky. We further restricted the final surface brightness profiles to be no 
more than 2.5 mag fainter than the estimated sky to guard against 
an artificial drop in the profile due to background contamination by galaxy light.
This procedure was used for all galaxies except NGC~2832, which extends well 
beyond the WFPC2 field even in the $B$ band. In this case, the estimate of 
$S_B$ would be badly biased so we fixed the $B$-band sky value by matching 
our photometry to the $B$-band profile of Peletier et~al. (1990).
The main advantage of this sky estimation method is that it guarantees
\textit{a priori} that the measured sky color will be close to its actual value. 

\begin{figure}
\plotone{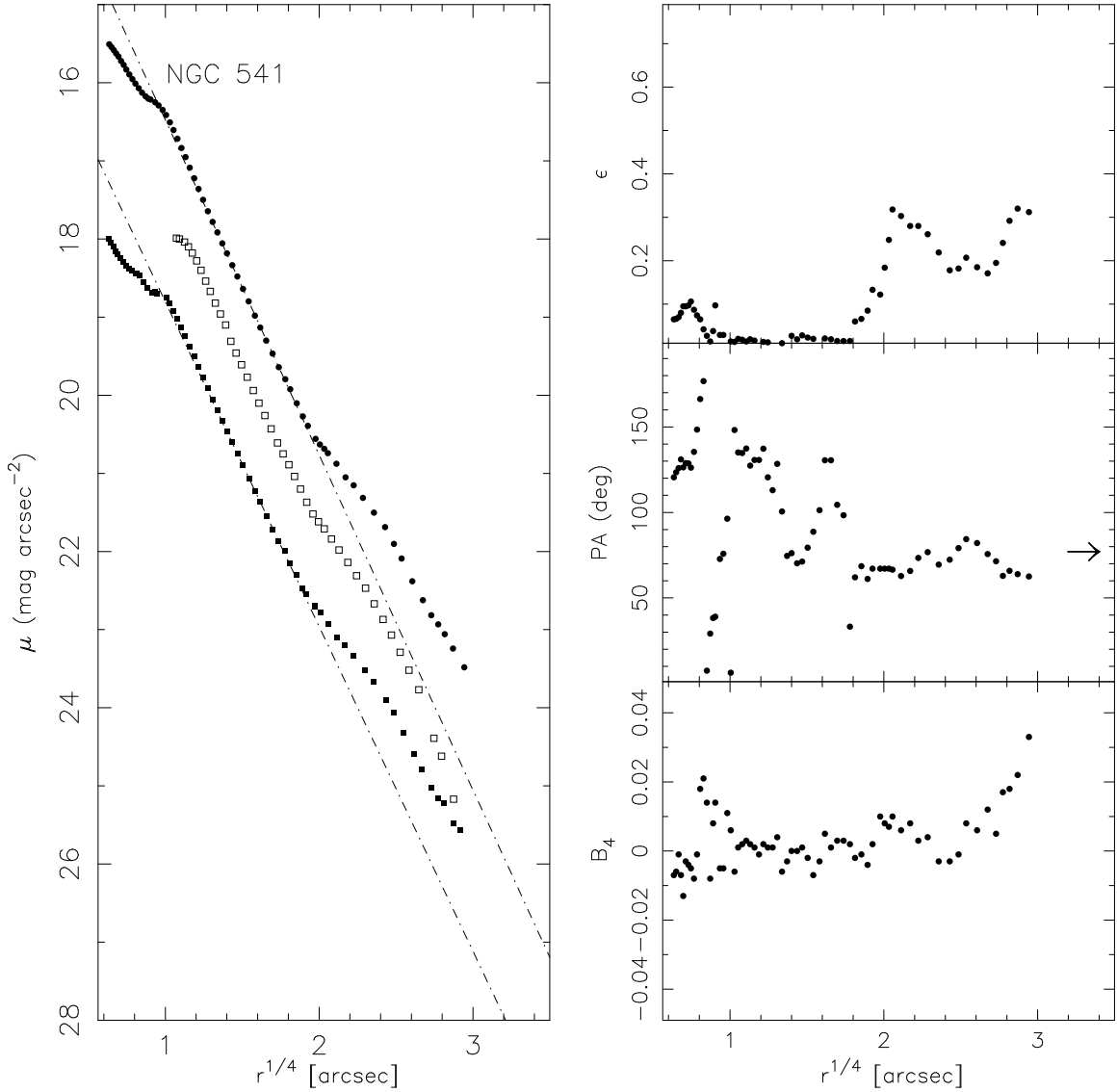}
\figcaption[Jordan.fig1.ps]{\textit{(Left Panel)}
Surface brightness profiles for NGC~541 in $B$ (filled squares) and $I$ (filled circles). 
The squares are a Gunn-Thuan $r$-band profile from de Juan et~al. (1994). 
The dot-dashed lines are best-fit de Vaucouleurs laws in the range $1.4 < r^{1/4} < 1.8$.
Note the flattening inside $r \simeq 1^{\prime\prime}$ caused by the
central dust disk (see Figure~\ref{fig:nuc}).
\textit{(Right Panels)} Ellipticity ($top$), position angle ($middle$) and $B_4$ ($bottom$) 
as a function of galactocentric distance. The arrow in the middle panel indicates the position 
angle of the major axis of Abell 194 (Lambas, Groth \& Peebles 1988; Plionis 1994).
\label{fig:n541sb}
}
\end{figure}

\begin{figure}
\plotone{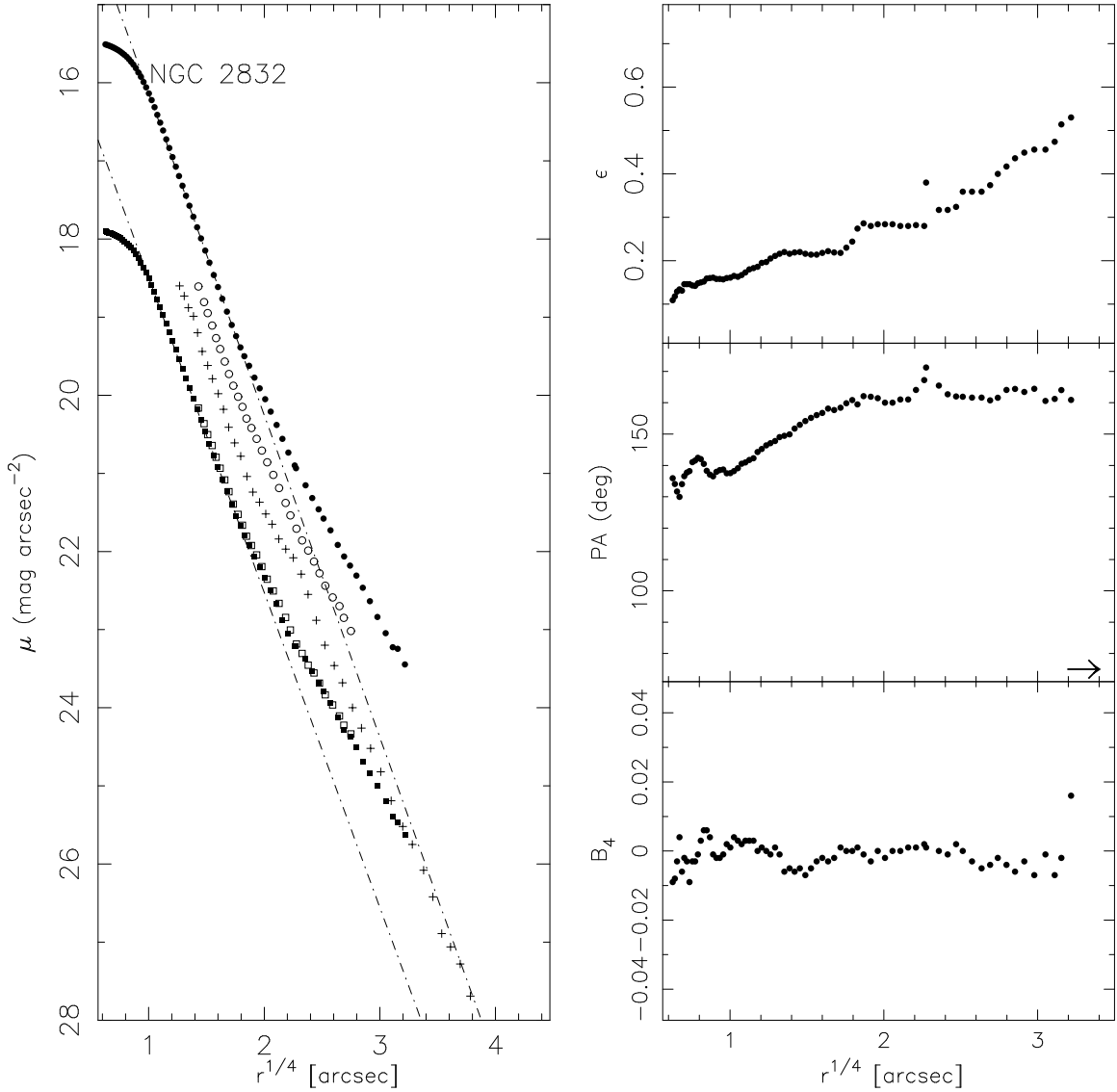}
\figcaption[Jordan.fig2.ps]{\textit{(Left Panel)}
Surface brightness profiles for NGC~2832 in $B$ (filled squares) and $I$ (filled circles). 
The squares and circles are respective $B$- and $R$-band profiles from Peletier et~al. 
(1990). Crosses denote the photographic profile of Schombert (1986).
The dot-dashed lines are best-fit de Vaucouleurs laws in the range $1.4 < r^{1/4} < 1.8$.
\textit{(Right Panels)} Ellipticity ($top$), position angle ($middle$) and $B_4$ ($bottom$) 
as a function of galactocentric distance. The arrow in the middle panel indicates the position 
angle of the major axis of Abell 779 (Plionis 1994).
\label{fig:n2832sb}
}
\end{figure}

Surface brightness measurements for the four galaxies are presented in
Table~\ref{tab:sb_all}, along with the results of our isophotal analysis.
Surface brightness profiles are shown in the leftmost panels of 
Figures~\ref{fig:n541sb}--\ref{fig:n7768sb}. Profiles 
from the literature are shown for comparison. We stress that, in the case of 
NGC~2832, the agreement at faint levels is forced upon our photometry for the
purpose of sky determination. For the remaining galaxies, the profiles are 
fully independent from those in the literature. In each case, the agreement is 
good except for the outer parts of NGC~541, where the profile of de Juan, 
Colina, \& P\'erez-Fournon (1994) turns downward compared to ours. We speculate 
that this downturn might be a consequence of their overestimating the 
background level: since our sky estimates are taken directly from the WFPC2 
images, they are unlikely to be underestimated.

Radial profiles of ellipticity, major-axis position angle and the 
$B_4$ parameter\footnote{$B_4$ indicates ``boxy'' isophotes if negative
and ``disky'' ones if positive (Jedrzejewski 1987).}
are shown in the right panels of Figures~\ref{fig:n541sb}--\ref{fig:n7768sb}.
As is common among BCGs, the ellipticities of NGC~541, NGC~4839 and NGC~7768 tend to 
increase with galactocentric radius ($e.g.$, Porter et~al. 1991) while
their major-axis position angles tend to align with those of the 
surrounding clusters ($e.g.$, Plionis 1994; West 1994; Kim et~al. 2002).
The ellipticity of NGC~2832 also increases with galactocentric radius,
but in this case, the galaxy's major-axis at large 
radii is oriented almost perpendicularly to the surrounding cluster.
Note that the profiles of NGC~541 and NGC~7768 flatten abruptly
at small radii ($r\lae 1''$). As Figure~\ref{fig:nuc} clearly shows,
the flattening in both case is due to the presence of compact dust disks in 
their nuclei. Tran et~al. (2001) find such compact disks in $\sim$ 18\% 
of the objects in their WFPC2 survey of early-type galaxies. It
is curious that two of the four cD galaxies in our sample contain
such disks.

\subsubsection{NGC~541: Comparison with VLT ground based photometry}
\label{sec:vlthst}

As mentioned in \S \ref{sec:intro}, the cD classification requires the 
presence of an envelope that departs from an $r^{1/4}$-law at low 
surface brightness. Our {\sl HST} profile for NGC~541 shows such an envelope 
starting at $r_b \sim 10^{\prime\prime}$, but we note that this galaxy 
has in the past been classified as both cD (Bird 1994) and D (Dressler 1980). 
To better characterize the surface brightness profile at large radii,
we obtained $B$ and $I$ images of NGC~541 with FORS1
on the VLT on 12 December, 2001 (see Figure~\ref{fig:n541vlt}).
Exposure times were $180$ and $20$ seconds in $I$, and $420$ and 
$60$ seconds in $B$. The short exposures were used to replace those pixels close
to the galaxy center that were saturated on the long exposures. The 
FWHM of isolated stars on the images was measured to be 
$\sim 1.1^{\prime\prime}$ in $B$ and $\sim 1.4^{\prime\prime}$ in $I$. 
The ground-based data, which have much greater areal coverage
than the {\sl HST} data and thus offer a more
direct means of measuring the local background level,
were reduced using standard IRAF routines and surface photometry
was performed with the task ELLIPSE. Because the observations were taken in 
non-photometric conditions, the {\sl HST} photometry was used to set the 
photometric zeropoints by matching the profiles in the range
$4^{\prime\prime} \lae r \lae10^{\prime\prime}$. 

The {\sl HST} and VLT profiles are compared in Figure~\ref{fig:vlthst}. An
envelope is clearly seen in the ground-based profile, confirming the
classification of NGC~541 as a cD galaxy. There is generally good agreement
between the two datasets, except at the faintest
levels where the {\sl HST} profile shows a steeper decline.
The increase in ellipticity for $r \gae r_b$ 
that was suggested by the {\it HST} data in Figure~\ref{fig:n541sb} is
readily apparent in the VLT data. It is also apparent that, in the transition
region between the PC and WF chips (typically around $R \sim 25^{\prime\prime}$),
the isophotal parameters derived from the {\it HST} data are somewhat less 
reliable because of the limited and varying areal coverage in this region.

As the ground-based surface brightness profile becomes shallower at large 
radii, $B_4$ increases dramatically. This behavior can be understood by
inspecting Figure~\ref{fig:n541vlt}. Beyond $r \sim 1^{\prime}$,
there is significant contamination by close companions, both to
the northeast and southwest. In fact, the various profiles overlap, 
producing a rather linear structure that runs diagonally through the 
field. As a result, the isophotes of NGC~541 have elongated ends that are 
aligned with this feature. This behavior, which is manifested as an increase in the
$B_4$ parameter for the outermost isophotes, is unlikely to be
is an artifact the sky estimation, since the slope of the surface brightness
profile changes rather abruptly, and setting the sky to obvious overestimates
does not make the increase in $B_4$ disappear. Given the presence
of significant structure in the background light, what we are seeing is
probably a consequence of our using a single value to characterize a
varying background in the vicinity of NGC~541. The surface brightness 
profiles shown in Figure~\ref{fig:vlthst} should therefore be treated
with some caution beyond $r \sim 1^{\prime}$.

\subsection{Point Sources: Detection, Classification and Photometry}
\label{sec:point}

GCs at the distances of
our program galaxies appear as unresolved point sources, even in the PC frames,
so their analysis is readily amenable to PSF-fitting photometry.
The images were first corrected for geometric distortion (Holtzman et al. 1995),
and detection frames were created by applying a ring median filter (Secker 1995)
with a radius of five pixels for the PC and four pixels for the WF frames in order
to eliminate the underlying galaxy light and to detect those sources
with scale-lengths smaller than the filter radius.
Detections were performed independently in both filters using SExtractor 
(Bertin \& Arnouts 1996) after setting the detection
threshold to three connected pixels with counts $2 \sigma$ above the local 
background in the detection image (which was convolved with a gaussian filter).
For some of our frames, there was an insufficient number of suitable stars
to build an PSF directly, so we created empirical PSFs from
archival observations and used these PSFs in the analysis of our program 
images.\footnote{The F814W and F450W PSF were constructed from images of the 
Sagittarius dwarf galaxy (GO6701), and Andromeda VI (GO8272), respectively.} 
Approximately one hundred stars per filter and chip were used to construct quadratically
varying PSFs with radii of ten pixels for the PC and five pixels for the WF chips. 
Using these PSFs, photometry was performed with DAOPHOT II (Stetson 1987; 1993), 
and the object lists in both the F814W and F450W filters were matched. Only 
those sources detected in both filters were retained.

Corrections for charge transfer efficiency were carried out using the prescriptions 
of Dolphin (2000), and standard {\sl HST} magnitudes were measured using the 
zeropoints given in Holtzman et~al. (1995).  The conversion to Johnson 
$B$ and $I$ was done using the following transformations (J. Holtzman, 
private communication), where $(B-I)_{1.3}\equiv (B-I)-1.3$,

\begin{widetext}
\begin{eqnarray*}
I = \left\{ \begin{array}{c} 
		m_{F814W}-0.025(B-I)_{1.3}-0.009(B-I)_{1.3}^2-0.021 \,\,\,\, \mbox{ if } \,\,\,\, (B-I)_{1.3}<0  \\
	  	m_{F814W}-0.012(B-I)_{1.3}-0.003(B-I)_{1.3}^2-0.021 \,\,\,\, \mbox{ if } \,\,\,\, (B-I)_{1.3}>0
	    \end{array} \right. \\
B=\left\{ \begin{array}{c}
	m_{F450W}+0.087(B-I)_{1.3}-0.008(B-I)_{1.3}^2+0.126 \,\,\,\, \mbox{ if } \,\,\,\,  (B-I)_{1.3}<0\\
	m_{F450W}+0.193(B-I)_{1.3}-0.047(B-I)_{1.3}^2+0.126 \,\,\,\, \mbox{ if } \,\,\,\, (B-I)_{1.3}>0
	\end{array} \right.
\end{eqnarray*}
\end{widetext}

A reddening correction was applied using the values of $E(B-V)$ obtained from the 
DIRBE maps of Schlegel, Finkbeiner \& Davis (1998) and the extinction curve 
of Cardelli, Clayton \& Mathis (1989). Additionally, a k-correction was applied to the
colors as described in \S\ref{sec:kcorr}. Since at this stage the candidate lists 
still contain significant contamination from compact background galaxies and foreground 
stars, additional selection criteria were imposed to obtain ``cleaned" lists of GCs. 
Specifically, we discarded objects whose DAOPHOT $\chi$ 
statistic\footnote{The DAOPHOT $\chi$ is a robust estimate of the ratio of 
the observed to expected
scatter about the model profile (Stetson \& Harris 1988).}
exceeded $\chi = 2$ in both $B$ and $I$, and 
whose SHARP parameters\footnote{The DAOPHOT SHARP parameter is an 
image radius index which is greater than zero if 
the object is more extended than the expected stellar profile and less than zero when the
detection appears sharper. The expected value for a point source is 
zero (Stetson \& Harris 1988).}
fell outside the range $-0.35 \lae$ SHARP $\lae 0.35$. We also discarded
objects with magnitudes more than $3\sigma$ brighter than the peak of the 
GC luminosity function (GCLF) in either $B$ or I, where $\sigma \sim 1.4$ mag 
is the typical dispersion of the GCLF (Harris 2001).  In doing so, we adopted
$M_B = -6.73$ and $M_I = -8.39$ mag for the GCLF ``turnover" magnitudes, 
displaced to a distance obtained from the mean cluster redshift using 
$H_0 = 72$ km s$^{-1}$ Mpc$^{-1}$ (Freedman et~al. 2001).

Artificial star tests were performed to calculate the completeness function and to assess
the extent to which our culling procedures would reject bonafide GCs. A total of 10,000 
stars were added to each chip, 50 stars each time so as not to alter significantly the crowding 
characteristics of our frames. Artificial stars had $B$ magnitudes in the range
$24 \lae B \lae 27$ mag and $B-I$ was randomly assigned a value assuming the color distribution
is a Gaussian with the observed mean and dispersion for each galaxy.
The artificial frames were then subjected to 
the same reduction procedure as the real data. Based on these tests, we conclude
that the above selection criteria reject $\lesssim 3\%$ of the artificial stars. 
Photometry for GC candidates in our four galaxies that have photometric
uncertainties less than $0.3$ mag in both $B$ and $I$ is presented in
Tables~\ref{tab:n541_gcphot}~--~\ref{tab:n7768_gcphot}.

\subsection{k-corrections}
\label{sec:kcorr}

Our program galaxies have redshifts in the range $0.018 \lae z \lae 0.027$. As we 
now show, redshift effects on the observed magnitudes and 
colors become significant at these distances. Since these effects depend rather strongly on
the object spectrum, we derive below k-corrections for both the
GCs and the host galaxies. In doing so, we make the assumption that the
GCs are old objects, meaning that they are roughly coeval with GCs in the
Milky Way. This assumption is supported by most observations of GCs
belonging to luminous early-type galaxies in cluster environments (see 
Jord\'an et~al. 2002 and references therein). Their broadband colors
are then determined mainly by metallicity, and not by age (Worthey 1994).

The k-correction is defined as
$$K(\lambda) = 2.5\log (1+z) + 2.5\log \biggl [ \frac{\int_0^{\infty}E(\lambda,t_0)S(\lambda)\,d\lambda}{\int_0^{\infty}
E(\frac{\lambda}{(1+z)},t_0)S(\lambda)\,d\lambda} \biggr ] \eqno{(4)}$$
where, following the notation of Poggianti (1997), $E(\lambda,t_0)$ is the luminosity
measured at wavelength $\lambda$ at the present time $t_0$ in the source rest frame,
and $S(\lambda)$ is the filter transmission curve.
In calculating k-corrections for the GCs, we used four template GC spectra, based
on observations of Galactic GCs, having 
different metallicities and different levels of ultraviolet emission\footnote{The results for 
cluster types with different ultraviolet behavior are very similar so the corrections were 
averaged.} (Bonatto, Bica \& Alloin 1995).  Specifically, we used their types
G2b/r, G3b/r, G4b/r, G5, which have respective
mean metallicities of [Fe/H] $=-0.4,-1.0,-1.5,-1.9$~dex.
To calculate the correction needed for solar-metallicity GCs, and for the galaxies
themselves, we used the elliptical template of Kinney et~al. (1996), which is based 
on spectroscopy of NGC~1399, NGC~1404, NGC~6868 and NGC~7196. The resulting k-corrections 
are presented in Table~\ref{tab:kcorr}.

Figure~\ref{fig:kcorr} shows k-corrections, $K(B-I)_0$, plotted as a function
of rest frame $(B-I)_0$. The upper and lower sequences show the corrections
for NGC~541 and NGC~7768, the galaxies with the lowest and highest redshifts in
our sample.
It is clear that the variation with $(B-I)_0$ is significant and must be accounted for.
We take the corrected colors to be linearly related to the observed $(B-I)_{\rm 0}$ by
$$(B-I)_{\rm 0,k} \equiv (B-I)_{\rm 0} - K(B-I)_{\rm 0} = C_K(B-I)_{\rm 0} + D_K \eqno{(5)}$$
where the constants are given in Table~\ref{tab:lincons}.

In principle, we might also wish to correct the observed colors for evolutionary effects.
However, any correction for the fading stellar populations 
in the color of these objects will be small compared to the k-correction. For instance,
using the populations synthesis models of Maraston (1998),
we find that the expected evolution in $(B-I)_0$ between $13$ and $14$ Gyr is 
${\delta}(B-I)_0 < 0.03$ mag for all metallicities. As the lookback time to our targets is
$\lesssim 0.3$ Gyr, the evolution in color is expected to be $\delta(B-I)\lesssim 0.01$ mag
--- significantly
smaller than the k-correction on the color. Note that the evolutionary effects 
on the observed magnitudes are, in general, \textit{not} negligible compared to the 
k-corrections for our targets. However, there is a partial cancellation in the corrections, 
as both $B$ and $I$ fade with age (albeit at different rates). For 
the present work, we have applied a k-correction to just the colors, since
we use the magnitudes only when calculating the specific frequencies in \S\ref{sec:sn}, and then 
only the $I$ band where the k-correction and evolutionary correction go in opposite 
directions. Thus, our results for the GC specific frequencies of these galaxies are 
quite insensitive to the correction ($i.e.,$ the $I$-band magnitudes vary by 
$\lesssim 0.03$ over the redshift range of our program objects).

\section{Analysis}

\subsection{Color Magnitude Diagrams and Color Distributions}

A remarkable finding from the past decade of GC research is that a majority of 
giant elliptical galaxies have GC systems with bimodal color distribution 
functions ($e.g.$, Gebhardt \& Kissler-Patig 1999; Larsen et~al. 2001). Because 
the colors of old stellar populations are more sensitive to metallicity than 
age, this bimodality implies the presence of two subpopulations of 
differing metallicity. 
In this section, we examine the color distributions of the GC systems
associated with our cD galaxies, thus probing the GC chemical enrichment
histories of early-type galaxies in their most extreme manifestation
of luminosity and size.

Figure~\ref{fig:cmd} presents color magnitude diagrams for the four GC systems.
All objects classified as unresolved by our culling procedure 
are plotted in this figure. Magnitudes and colors have been corrected for 
reddening and extinction, and k-corrections have been applied to the observed colors.
It is apparent that none of the GC systems show obvious sequences arising from 
the presence of two subpopulations. This impression is confirmed in 
Figure~\ref{fig:hists}, where we plot color distributions for all objects 
with $\sigma({B-I}) \leq 0.3$ mag. The solid line is a kernel 
density estimate of the color distribution (Silverman 1986) obtained with 
a normal kernel. The adopted smoothing parameter $h$ was chosen using 
the prescription of Silverman (1986) which is well suited to a wide range in
densities and is trivial to evaluate: $h=0.9An^{-1/5}$ 
where $A=min(\mbox{standard deviation},\mbox{interquartile range}/1.34)$. 
The dashed curves show $90\%$ confidence bands obtained with a 
smoothed bootstrap procedure (Silverman 1986).
 
We have computed a variety of statistical indices for each of the color distributions 
shown in Figure~\ref{fig:hists}, including biweight estimates of location, $C_{BI}$,
and scale, $S_{BI}$, Finch's (1977) asymetry index, $AI$, (akin to skewness) and 
Hoaglin, Mosteller \& Tukey (1983) tail index, $TI$, (akin to kurtosis).\footnote{For
a description of the asymetry and tail indices and some of their properties, see Bird 
and Beers (1993). See Beers, Flynn \& Gebhardt (1990) for a description of the 
biweight estimators.} The results 
are summarized in Table~\ref{tab:stats}, where the upper and lower limits on each 
index represent $68\%$ confidence bands obtained with a bootstrap procedure.
To within the uncertainties, the width of the distributions
is the same for all galaxies, and it
is clear that none of the distributions show statisitically significant 
skewness or extended tails beyond what would be expected from a parent Gaussian
distribution. Similarly, the DIP test (see Gebhardt \& Kissler-Patig 1999) accepts 
the hypothesis of unimodality for each of the distributions shown in 
Figure~\ref{fig:hists}. We caution, however, that although bimodality is formally 
rejected in each case, the photometric errors, 
$\sigma{(B-I)} \sim 0.2$ mag, are sufficiently large that intrinsic
bimodality (with an expected separation between the modes of 0.3$-$0.4 mag in
$B-I$; $e.g.$, Forbes et~al. 1998) would be difficult to detect given 
the current samples.

\subsection{Color Profiles and Offsets}

Mean colors for our program galaxies and their GC systems are given in 
Table~\ref{tab:colorinfo}, which summarizes various observed and derived photometric data.
From left to right, this table gives the galaxy name, adopted reddening, average colors
for the galaxy and GCs, the total $V$-band magnitude from RC3, the galaxy redshift, 
its total $V$-band absolute magnitude, and approximate luminosity in units of $L^*$.
In Figure~\ref{fig:bivsr}, we plot the colors of
the GC candidates as a function of galactocentric distance for each of the four galaxies. 
The solid line in each panel is the best-fit linear relation for GCs with 
$\sigma({B-I}) \leq0.25$ mag. The dashed-lines are profiles of the galaxy light measured
from our {\sl HST} images. In all four cases, as is typical for GC systems, there is a 
negative color gradient in the sense that the GC system becomes bluer with increasing 
galactocentric distance ($e.g.$, Strom et~al. 1981).
The GC color gradients parallel those of the underlying galaxies, with mean offsets
ranging from 0.33~mag for NGC~541 to 0.50~mag for NGC~7768. If we make the somewhat
daring assumption that the composite stellar populations of the galaxies follow the 
same color-metallicity relation as the globulars (Barmby et~al. 2000), then these 
differences correspond to metallicity offsets in the range 
$0.8 \lae {\Delta}{\rm [Fe/H]} \lae 1.3$~dex.

Indeed, the color/metallicity offset apparent in Figure~\ref{fig:bivsr} is a ubiquitous feature 
that is observed in galaxies ranging from faint, low-surface dSph/dE galaxies to the 
brightest and most massive early-type galaxies.  The left panel of 
Figure~\ref{fig:metoffset} shows mean GC and stellar metallicity plotted as a function
of host galaxy magnitude for a large sample of early-type galaxies. Note that this 
plot is an updated version of Figure~1 of C\^ot\'e et~al. (2000), to which we have 
added elliptical galaxies from the sample
from Kundu \& Whitmore (2001) and our four cD galaxies (shown as the circled
points). Improved stellar metallicity determinations for several dwarf galaxies have 
been taken from Shetrone, C\^ot\'e \& Sargent (2001) and Tolstoy et~al. (2003).
For those galaxies from the survey of Kundu \& Whitmore (2001), GC metallicities 
were calculated by combining their $(V-I)$ colors with the $(V-I)$-metallicity relation
of Barmby et~al. (2000). Metallicities for the galaxies themselves come
from the integrated-light spectroscopy of Trager et~al. (2000) when available; for
those galaxies lacking spectroscopic data, we estimated crude metallicities
by combining their $(B-V)$ colors from RC3 (de Vaucouleurs et~al. 1995)
with the $(B-V)$-metallicity relation of Barmby et~al. (2000).

It is clear from the left panel of Figure~\ref{fig:metoffset} that the GCs and stars obey 
reasonably tight relations that run roughly parallel to each other. The right panel
of this figure shows, for those galaxies that have metallicity determinations for both 
the stars and the GC systems, the offset between the two components. 
Remarkably, this offset is nearly constant over the whole range of galaxy magnitude.
An $F$-test indicates that the offset appears to be constant over the full
range in galaxy magnitude, with a weighted mean value of $\Delta$[Fe/H]$=0.79\pm0.04$~dex.
We conclude that the metallicity offset between stars and GCs in these
luminous cD galaxies is indistinguishable from that observed in
normal ellipticals. This metallicity offset presents 
another apparently ``universal" property of GC systems --- analogous to the 
near-Gaussian nature of the GC luminosity function (Harris 2001) and the 
apparently constant efficiency of GC formation (Blakeslee et~al. 1997; McLaughlin 1999a) --- 
that must be explained by any viable model of GC formation.

\subsection{Specific Frequencies}
\label{sec:sn}

The GC specific frequency, $S_N=N_{\rm GC}10^{0.4(M_V+15)}$, is a convenient measure of 
GC formation efficiency relative to field stars in the underlying galaxy (Harris \& 
van den Bergh 1981; cf. McLaughlin 1999a). Possible dependencies of $S_N$ on 
morphological types have played an important role in the debate over 
how, and when, early-type galaxies form (Schweizer 1987; Ashman \& Zepf 1992;
van den Bergh 1995).
Since the first ``high-$S_N$" giant galaxies to
be identified (M87, NGC~1399, NGC~3311) were all cDs, it was believed
that the cD environment was somehow conducive to the efficient formation
of GCs relative to field stars (Harris 1988). However, with the acquisition of 
more and better data,
it became apparent that not all cD galaxies had high specific frequencies
($e.g.$, Harris, Pritchet \& McClure 1995), leading West et~al. (1995) to propose a
phenomenological model in which the GC specific frequency of BCGs correlates
with the mass of the surrounding galaxy cluster. This suggestion was later
confirmed observationally as a correlation between specific frequency and 
cluster velocity dispersion or X-ray temperature (Blakeslee, Tonry \& Metzger 
1997; Blakeslee 1999). In other words, it appears that the properties of the 
surrounding galaxy cluster are connected to both the cD envelope {\sl and} the 
overall number of GCs, and that the mere presence of an envelope is not enough 
to guarantee a high GC specific frequency.

To measure GC specific frequencies for our program galaxies, we must first
estimate the sizes of their GC systems using the bright end of the GCLF, which we
observe directly. For data such as ours, which do not reach the turnover,
any attempt to fit simultaneously for the number of GCs, 
the turnover magnitude, and the dispersion of the GCLF 
would be confounded by large and highly correlated errors
in the derived parameters (Hanes \& Whittaker 1987; Secker \& Harris 1993).
We therefore proceed by fixing the dispersion of the GCLF at $\sigma = 1.4$~mag and 
adopting $M_{\rm TO} = -7.33$~mag for the $V$-band absolute magnitude 
its turnover (Harris 2001).

To obtain the completeness-corrected luminosity function, we divide our 
GC sample into three radial regions: 
$r\leq 20'',\, 20''<r\leq 50''$ and $r>50''$ .
We determine the completeness function, $f(m)$, in each radial bin by fitting the function 
$$f(m)=\frac{1}{2}\left[ 1-\frac{\alpha (m-m_{lim})}{\sqrt{1+\alpha^2(m-m_{lim})^2}} \right] \eqno{(6)}$$
to the results of our artificial star tests, where $m_{lim}$ is the magnitude 
at which $f=0.5$ (Fleming et~al. 1995). These functions were then used to correct 
the luminosity functions in each bin up to a magnitude obtained by averaging $m_{lim}$ 
for the three regions.  The three LFs were then added to find the completeness 
corrected LF for the full field. 

We fit for $N_{\rm GC}$ in a GCLF of the form 
$$N(m) = N_{\rm GC}(2\pi\sigma^2)^{-1/2}\exp[-(m-m_{\rm TO})^2/(2\sigma^2)] \eqno{(7)}$$
where $m_{\rm TO} = (m-M) + M_{\rm TO}$. Here $M_{\rm TO}$ is absolute magnitude
of the GCLF turnover and $(m-M)$ is the 
distance modulus obtained from the observed redshift (see
Table~\ref{tab:prop_clus}) and $H_0 = 72\pm8$ km~s$^{-1}$~Mpc$^{-1}$ (Freedman et~al. 2001),
which is consistent with the value obtained from WMAP (Spergel et~al. 2003). 
We assume the uncertainty in the turnover magnitude is due 
to the uncertainty of $0.05$ quoted in Harris (2001) plus the 
joint uncertainties arising from the distance 
determination (as a result of deviations from a pure Hubble flow) and
the uncertainty in the Hubble constant itself. 
The combined error in the turnover magnitude due to the latter effects is 
$$\sigma^2_{m}=(5/ln(10))^2(\sigma_{cz}/cz)^2+(5/H_0ln(10))^2\sigma^2_{H_0} \eqno{(8)}$$ 
where we have adopted a constant value of $\sigma_{cz} = 300$ km s$^{-1}$ for the 
$rms$ cluster peculiar velocity (Bahcall \& Oh 1996). Clearly, these uncertainties 
should be viewed as lower limits since background contamination, photometric errors 
for the GCs, and model assumptions have not been included.

We take $N_{\rm GC}$ in equation (7) as our best estimate for the
total number of GCs within our field. Since this is a {\it metric}
quantity, it is denoted as $N_{\rm GC}^{\rm met}$ hereafter. For each
galaxy, $S_N^{\rm met}$ is measured by calculating the integrated
magnitude of the underlying galaxy within our WFPC2 field with the aid
of the $B$- and $I$-band galaxy models presented in \S~\ref{sec:sb}. To convert
these $B$- and $I$-band magnitudes into the $V$ bandpass, we assume typical colors
for ellipticals as given in Fukugita et~al (1995): $\langle B-V\rangle = 0.96$ mag and
$\langle V-I\rangle = 1.31$ mag.
Table~\ref{tab:snval} summarizes the GC specific frequency measurements for our
cD galaxies. From left to right, this table records the galaxy name, 
absolute magnitude, total number of globular clusters, 
and GC specific frequency. Once again, these are 
all {\it metric} quantities. 

It is interesting to note that 
none of these galaxies show high specific frequencies,
their values being typical for ellipticals and even somewhat low for NGC~7768. 
Blakeslee et~al. (1997) measured specific frequencies for NGC~2832, NGC~4839
and NGC~7768 as part of their survey of BCGs. Within the errors, our results are consistent 
with theirs for all three objects. NGC~7768 was also studied by Harris, Pritchet \& McClure (1995), 
who obtained $S_N=3\pm2$. Their result scales with their assumed value of $S_N = 15$ 
for M87. In light of more recent measurements the GCLF parameters in M87 by Whitmore et~al. (1995),
we adopt $S_N = 11$ for this galaxy (using the correction factor presented in equation 17 of
McLaughlin, Harris \& Hanes 1994). The Harris et~al. (1995) measurement
for NGC~7768 then falls to $S_N \sim 2.3$. This is in good agreement with 
our determination of $S_N = 2.2 \pm 0.8$. Finally, Mar\'{\i}n-Franch \& Aparicio (2002)
find $S_N=7.0\pm1.9$ for NGC~4839 --- slightly higher than, but still 
consistent with, our measurement of $S_N=5.6\pm1.9$.
This is the first determination of $S_N$ for NGC~541 and its value --- like those
of the two other galaxies for which X-ray luminosities of the host cluster are
available (see Table~\ref{tab:prop_clus}) --- is consistent with that expected from the cluster's 
X-ray temperature and the $S_N$-$L_X$ correlation presented by Blakeslee et~al. (1997).

\section{Constraints on the Merger Histories of cD Galaxies}

As discussed in \S\ref{sec:intro}, the physical properties of cD galaxies, coupled
with their unique locations near the dynamical centers of rich galaxy clusters, 
suggest that some aspects of their formation and/or evolutionary histories 
differ from those of normal elliptical galaxies. A wide range of evolutionary 
phenomena has been invoked to explain their properties, including
cannibalism, tidal stripping, and star formation in coolings flows. Clearly, a 
resolution of the debate concerning which, if any, of these processes has played a 
role in the formation of cD galaxies would benefit from a knowledge of their
merger histories. 

\subsection{Protogalactic Mass/Luminosity Spectra From Globular Cluster Metallicity Distributions}
\label{sec:pmf}

A method of using GC metallicity distributions to determine protogalactic mass 
spectra has been described by C\^ot\'e et~al. (1998; 2000; 2002). A full description 
of the method is presented in these papers, so here we simply note its basic 
assumptions and limitations. In brief, the method assumes that the protogalactic 
luminosity (and mass) spectrum can be 
approximated by a Schechter function (Schechter 1976), and that the present-day GC 
specific frequency of any surviving protogalactic ``fragments" is $S_N \approx 4$.
It is further assumed the galaxy under consideration has been assembled hierarchically,
in a primarily dissipationless manner, so that no GCs are formed during mergers
and interactions. In other words, star and GC formation during mergers 
are ignored entirely in 
this approach. The metallicities of individual GCs are instead taken to reflect
differences in the depths of the gravitational potential wells in which they 
formed. Indeed, the tight empirical relation between mean GC metallicity and total 
galaxy magnitude, such as that shown in Figure~\ref{fig:metoffset}, is 
{\it prima facie} evidence for the importance of local environment in the chemical 
enrichment of GC systems. As a consequence, we adopt a GC metallicity-host galaxy 
luminosity relation that is based on nearby dwarf galaxies and the metal-rich
(bulge) GC populations of M31 and the Milky Way (C\^ot\'e et~al. 2002).

The implementation of the model proceeds as follows. Using the above model inputs, simulated
GC metallicity distributions are generated over a grid in two parameters
chosen to describe the mass spectrum\footnote{In converting to masses, we assume
a constant mass-to-light ratio of $\Upsilon_V = 5$ for all protogalactic fragments.} 
of protogalactic fragments from which the galaxy was assembled: (1) the slope,
$\alpha$, of the Schechter function from which the simulated protogalactic fragments 
were drawn; and (2) the ratio of the mass of largest protogalactic fragment, 
${\cal M}^1$, to the final mass, ${\cal M}_{\it f}$ of the assembled galaxy:
$$\zeta= {\cal M}^1/{\cal M}_{\it f}. \eqno{(9)}$$
For each simulated GC system, we extract at random an identical
number of GCs as contained in the actual sample and
add the appropriate amount of measurement error to the simulated colors.
The transformation between color and metallicity is performed using the 
transformations of Barmby et~al. (2000). We restrict the sample of GCs 
to lie on the range 1.23 $\le (B-I)_{\rm 0,k} \le$ 2.49~mag, which is equivalent 
to $-2.5 \le$~[Fe/H]~$\le1.0$~dex. This generous range is expected to include the 
vast majority of bona fide GCs in these galaxies (see, $e.g.$, 
Beasley et~al. 2000; Cohen, Blakeslee \& C\^ot\'e 2003).
 
For each simulated GC system, a Kolmogorov-Smirnov (KS) test is performed to test 
the hypothesis that the simulated and observed metallicity distributions have 
been drawn from the same parent distribution.
Simulations are carried out $100$ times at each point in the $\alpha$--$\zeta$ plane,
giving a grand total of 60,000 simulations over the full grid. At each grid
element, the average KS probability is recorded. The resulting probability surface,
shown in Figures~\ref{fig:ks541},~\ref{fig:ks2832},~\ref{fig:ks4839} and
\ref{fig:ks7768}, gives the likelihood that the observed GC color/metallicity distributions 
can be reproduced with a protogalactic spectrum characterized by that 
$\alpha$--$\zeta$ pair. The four crosses in each figure mark representative simulations
for each galaxy that have a better than $99\%$ probability of being drawn from the same 
distribution as the observed GC metallicity distribution function. 
The upper panels of Figures~\ref{fig:mdf541},~\ref{fig:mdf2832},~\ref{fig:mdf4839} and
\ref{fig:mdf7768} show comparisons of the observed GC 
metallicity distribution functions with the results from these four simulations;
the lower panels of these figures show the corresponding protogalactic mass 
spectra.

The key result from these simulations 
is that the observed GC color/metallicity distributions seem to be
perfectly consistent with those expected if cD galaxies form through
the accretion of numerous galaxies and protogalactic fragments that
managed to form the bulk of the GCs before being integrated into the central galaxy.
At the same time, Figures~\ref{fig:ks541}--\ref{fig:mdf7768}
reveal that a wide variety of protogalactic mass spectra are
able to reproduce the observed metallicity distributions. For comparison,
C\^ot\'e et~al. (2002) carried out an analysis of the 28 early-type galaxies 
of Kundu \& Whitmore (2001) using the method described above, 
finding $\langle \zeta \rangle = 0.25\pm0.03$ (mean
error) and a rather steep slope of $\langle \alpha \rangle = -1.88\pm0.03$
(mean error). While these values are certainly consistent in all four cases
with those found here, Figures~\ref{fig:ks541}--\ref{fig:mdf7768}
show that, with the possible exception of NGC~4839, 
shallower slopes are also allowed. Aside from NGC~4839 (which we discuss in detail below),
the locus of most probable $\alpha$--$\zeta$ pairs for each galaxy tend to 
form an elongated ridge that stretches
from the lower-left corner of the plot ($i.e.,$ low $\zeta$ and shallow
slopes) to the middle-right region ($i.e.,$ intermediate $\zeta$ and steep
slopes). The weaker constraints on $\alpha$ and $\zeta$
compared to those found by C\^ot\'e et~al. (2002), are a direct
consequence of the fact that the GC metallicity
distributions for these cD galaxies are dominated by photometric
errors. With the current data, we are able to conclude that the GC 
metallicity distributions are consistent with the cDs
being assembled via mergers, accretion and tidal stripping, but we 
are unable to place firm constraints on the detailed shape of the mass 
spectra of the cannibalized objects (with the possible exception of 
NGC~4839; see below). We now turn out attention to the mechanism(s)
responsible for these putative mergers.

\subsection{Dynamical Friction and the Luminosity Function of Cannibalized Galaxies}

Note that above discussion refers implicitly to the {\it initial} luminosity
function of galaxies, protogalaxies and protogalactic ``fragments". Since 
our simulations treat mergers in a purely statistical manner, equal merger 
probabilities are assumed for all galaxies and protogalactic
fragments. This assumption is clearly an oversimplification: if 
mergers proceed through dynamical friction, the merger probability 
will be a function of mass ($e.g.$ Binney \& Tremaine 1987). In this section, 
we examine how dynamical friction is expected to modify the initial 
luminosity function. Our aim is
compare the luminosity function of cannibalized
galaxies predicted by dynamical friction arguments with that obtained in
our analysis of the GC metallicity distributions.  

At the centers of clusters where cD galaxies reside, the galaxy luminosity 
function will evolve with time because the rate of dynamical friction is mass
(and, hence, luminosity) dependent. If cannibalism as a result of dynamical 
friction is the primary mechanism by which cD galaxies evolve, then we would 
expect the luminosity function of cannibalized galaxies to be skewed toward
luminous galaxies. Note that, for our purposes, we are not concerned with 
the {\it rate} at which galaxies sink to the cluster center, but merely
with the final luminosity distribution of all galaxies cannibalized over 
the cluster lifetime.

We assume that the initial luminosity function is given by a Schecter function,

$$\phi_i(L) = n_{*}(L/L_{*})^{\alpha_i}\exp(-L/L_{*})L_{*}^{-1}, \eqno{(10)}$$
and that the joint luminosity and spatial density distribution of galaxies within the 
cluster, $n(r,L)$ 
is given by the product of $\phi_i(L)$ with a spatial distribution $\eta(r)$, such that 
$n(r,L)=\phi_i(L)\eta(r)$. A galaxy orbiting on a circular orbit in an isothermal 
potential reaches the cluster center after a time (Binney \& Tremaine 1987)
$$t_{f} = \frac{1.65 r_i^2\sigma}{GM \ln{\Lambda}}. \eqno{(11)}$$
Here $r_i$ is the initial distance of the galaxy, $\sigma$ is the cluster velocity
dispersion, and $\ln{\Lambda}$ is the standard Coulomb logarithm. Setting $t_{f}$ 
to the age of the cluster, $T_{clus}$, we can solve for the distance, $r_i(M)$, within 
which galaxies of mass $M$ will have spiraled to the center. Assuming a constant 
mass-to-light ratio, $\Upsilon$, we can also express this distance as a function of $L$,

$$r_i(L)= \biggl [ \frac{{\Upsilon}T_{clus}G\ln{\Lambda}}{\sigma}\frac{L}{1.65}\biggr ]^{0.5} \propto L^{0.5} \eqno{(12)}$$

For the host clusters of our cD galaxies, $r_i$ varies between $\sim$~13 and 
$\sim$~1300~kpc for galaxy masses in the range 10$^8$ to 10$^{12}$~$M_{\odot}$. 
The luminosity function of cannibalized galaxies will
be given by

$${\phi}_c(L)=4\pi \phi_i(L) \int_0^{r_i(L)}\eta(r)r^2\,dr. \eqno{(13)} \label{eq:phi}$$

The surface density profile of galaxies in the inner regions of clusters goes as
$\rho(r) \propto r^{-1}$ ($e.g.$, Beers \& Tonry 1986; 
Merritt \& Tremblay 1994; McLaughlin 1999b). For a spherical cluster, this
corresponds to a three-dimensional density distribution of 
$\eta(r) \propto r^{-2}$, which in turn implies
$${\phi}_c(L) \propto \phi_i(L)L^{0.5}. \eqno{(14)}$$
In other words, the slope of the luminosity function of cannibalized galaxies, 
$\alpha_c$, is related that of the initial luminosity function, $\alpha_i$, 
through the relation 
$$\alpha_c \simeq \alpha_i + 0.5. \eqno{(15)}$$

This expression provides a a crude connection between the expected luminosity 
function of cannibalized galaxies to
that of the surrounding cluster at the time of its formation
under the simplifying assumptions made. 
Note that except in the case of a very steep initial density distribution,
$\alpha_c \gae \alpha_i$ regardless of the exact value of $\alpha_c-\alpha_i$.
 
\subsection{Comparing the Luminosity Functions}

Is it plausible that these cD galaxies were
assembled mainly through dynamical friction? We may examine this 
possibility by assuming, for the time being, that the protogalactic luminosity 
functions found in \S\ref{sec:pmf} are the outcome of cannibalism
through dynamical friction, so that $\alpha \simeq \alpha_c$ and 
$\alpha_i = \alpha - 0.5$.

For NGC~541, NGC~2832 and NGC~7768, 
Figures~\ref{fig:ks541}-\ref{fig:mdf2832} and 
\ref{fig:ks7768}-\ref{fig:mdf7768} 
show that a rather wide range of protogalactic
luminosity functions are capable of reproducing the observed GC
metallicity distributions in these galaxies. 
Depending on the
precises values of $\zeta$, slopes in the range $-2 \lae \alpha \lae -0.75$
are capable of yielding simulated metallicity distributions that are in close 
agreement with those observed; the constraints on $\alpha$ are somewhat 
tighter in the case of NGC~4839, with $-2 \lae \alpha \lae -1.1$.
These values of $\alpha$ would require the host
clusters ($i.e.$, Abell 194, Abell 779 and Abell 2666) to have
initial luminosity functions with slopes in the range
$-2.5 \lae \alpha_i \lae -1.25$. For Abell~1656, the corresponding limits
are $-2.5 \lae \alpha_i \lae -1.65$.
Are such slopes feasible, or can they
be ruled out from observations of the present-day luminosity functions in 
these clusters? 

There is, to the best of our knowledge, no published luminosity
function for Abell~779, so we restrict ourselves to Abell~194, Abell~1656 and 
Abell~2666. 
Interestingly, there have been recent reports of steep 
luminosity functions in both Abell~194 and Abell~2666.
Trentham (1997) finds $\alpha_{\rm obs} > -1.6$ in Abell~194, while de 
Propris et~al. (1995) claim $\alpha_{\rm obs} \sim -2.2$ for Abell~2666. 
We caution, however, that in both cases, the measured luminosity 
functions are uncertain to the point being essentially unconstrained
($e.g.$, Trentham quotes a best-fit value of
$\alpha_{\rm obs} = -2.2^{+\infty}_{-\infty}$ for Abell~194).
Firm conclusions on the viability of dynamical friction as the
main mechanism responsible for the formation of these
galaxies must await the measurement of improved GC metallicity
distributions and, equally important, accurate luminosity functions
for their host clusters.

Observational constraints on the luminosity function in 
Abell~1656 are of much higher quality. Several comprehensive studies of 
the luminosity function in this cluster which have yielded results that 
are in good agreement:
$\alpha_{\rm obs} = -1.42\pm0.05$ (Bernstein et~al. 1995),
$\alpha_{\rm obs} = -1.41\pm0.05$ (Secker, Harris \& Plummer 1997),
$\alpha_{\rm obs} = -1.33\pm0.06$ (Beijersbergen, Schaap \& van der Hulst 2002), and
$\alpha_{\rm obs} = -1.31 \pm0.05$ (Mobasher et~al. 2003).
Strictly speaking, these measurements refer to the {\it global} luminosity
function of the Abell~1656 cluster
and not the subcluster that hosts NGC~4839, but Mobasher et~al.
(2003) find the luminosity function in the NGC~4839 subcluster to
be indistinguishable from that of the cluster as a whole.
The average of the above values, $\langle \alpha_{\rm obs} \rangle = -1.37$,
falls outside the range of probable $\alpha_i$ values given above
($i.e.$, $-2.5 \lae \alpha_i \lae -1.65$).
We tentatively conclude that the hypothesis that cD galaxies 
were assembled mainly through mergers driven by dynamical friction
seems inconsistent with the available data for NGC~4839 and
Abell~1656, the galaxy/cluster pair in our sample with the highest
quality GC metallicity distribution and luminosity function.
Recall from \S\ref{sec:intro} that various dynamical evidence suggests
that cD galaxies typically grow by only $\approx$ 1-2$~L^*$ as a result of 
cannibalism through dynamical friction during the post-virialization regime
($e.g.$, Merritt 1985; Lauer 1985; Dubinski 1998).
Since NGC~4839 has a luminosity of $\sim 7 L*$, this conclusion is in 
agreement with the expectation from the dynamical evidence that
dynamical friction alone can not account for the observed luminosity.

\section{Conclusions}

We have nearly doubled the number of cD galaxies with available color/metallicity distributions
for their GC systems, thus probing the realm of early-type galaxies in their 
most extreme manifestation of luminosity and size. Despite the undeniably special 
nature of these galaxies, their GC systems seem remarkably normal in all the 
properties we have explored: specific frequency, mean metallicity, and the 
metallicity offset between the GCs and host galaxy stars. An analysis of the
protogalactic mass spectra for these galaxies, while more model
dependent, similarly reveals no obvious anomalies compared to normal ellipticals.

The GC specific frequencies of these cD galaxies fall within the expected 
range given the known scaling relations between specific frequency and cluster X-ray 
temperature and velocity dispersion. Our specific frequencies corroborate 
previous measurements for three of our galaxies; the measured specific frequency
for the remaining object, NGC~541, is consistent with expectations given the
mass of Abell~194, the surrounding cluster. Thus, our findings are
consistent with a picture in which the efficiency of GC formation scales with
the total mass of the cluster itself, implying that the GC formation efficiency per unit
total mass ($i.e.$, including dark matter) is universal (Blakeslee et~al. 1997; Blakeslee 1999).
It would be interesting to measure density profiles for the
X-ray emitting gas these clusters,
to see if the total numbers of GCs in these galaxies are also consistent with a 
constant efficiency of GC formation per {\it baryon} mass, as suggested by McLaughlin (1999a).
In any event, it seems clear that the GC systems belonging to these cD galaxies trace 
the cluster properties so that, as Blakeslee (1999) notes, it may be more 
correct to view ``high-$S_N$" galaxies as being underluminous given
their total masses, rather than ascribing to them an unusually high GC formation 
efficiency. Possible explanations for this ``missing light'' problem include tidal
heating of the gas as the galaxy cluster collapses (Blakeslee et~al. 1997), and
starbust driven galactic winds (Harris, Harris \& McLaughlin 1998)
which could be responsible for halting star formation in BCGs to an extent 
that is controlled by the cluster's mass.

Our results suggest that, whatever mechanism is responsible for the formation of
cD galaxies, perhaps making some of them appear underluminous compared to their
host clusters in the process, {\it it seems to leave the GCs unscathed}. 
Indeed, the metallicity offset between the CG system and the underlying
galaxy is indistinguishable from that observed for other early-type galaxies, and
thus presents us with another uniformity which must be
tied in a fundamental way to their formation process. A plausible origin for this 
offset may rely on the timing of GC formation. We find the mean metallicity
of the gas out of which GCs formed to be roughly six times lower than that of 
galaxy itself; if the GCs form before the bulk of the stars, then a metallicity 
offset seems a natural, if not inevitable, outcome. This admittedly qualitative 
scenario agrees with the idea that star formation was inhibited for some central
galaxies but not their GC systems, as the GCs might have formed prior to the onset of 
this as-yet-unidentified process.

Different models for the formation of cD galaxies differ mainly on the timescales over
which the cD galaxies form. We have examined the possibility that cD galaxies have
been assembled through the accretion of numerous galaxies and protogalactic fragments
that managed to form their GCs prior to being captured and disrupted
by the central galaxy. Our key finding in this regard is that such a process
appears to be entirely consistent with the observed GC metallcity distributions 
for these galaxies. Unfortunately, a wide variety of protogalactic mass spectra are
able to reproduce the observed metallicity distributions given the rather
rather poorly constrained metallicity distributions. 

What mechanism might be responsible for such mergers? In the case of NGC~4839,
the cD in the southwest extension of Abell~1656 and the galaxy in our
sample having the most reliable GC metallicity distribution, we find 
the present-day luminosity function of Abell~1656 to be too shallow to be consistent
with the assembly of the galaxy from dynamical friction. In other words, while 
it is possible to choose an ensemble of galaxies, protogalaxies and protogalactic 
fragments that, once assembled, would produce a GC system with the appropriate
metallicity distribution, the required luminosity function of these objects
seems overly steep compared to that measured at the present time.
This suggests that, in agreement with the previous conclusions
based on dynamical arguments (Merritt 1985; Lauer 1985), cannibalism 
through dynamical friction is not the primary means by which cD galaxies 
are assembled. The properties of the GC system instead favor a scenario 
($e.g.$, Dubinski 1990) in which the cDs form rapidly, and at early-times, 
via hierarchical merging prior to cluster virialization.

\acknowledgments

We thank J. Holtzman for providing the transformation equations used in 
\S\ref{sec:point}, and K. Gebhardt for providing routines to calculate the DIP statistic.
Support for program GO-8184 was provided by NASA through a grant from the Space Telescope 
Science Institute, which is operated by the Association of Universities for Research in 
Astronomy, Inc., under NASA contract NAS5-26555.
Support for this work was provided by the National Science Foundation through a grant 
from the Association of Universities for Research in Astronomy, Inc., under NSF 
cooperative agreement AST-9613615 and by Fundaci\'on Andes under project No.C-13442.
MJW acknowledges support from NSF grant AST-0205960.
DM is partially supported by FONDAP Center for Astrophysics 15010003.
This research has made use of the NASA/IPAC Extragalactic Database (NED) which is 
operated by the Jet Propulsion Laboratory, California Institute of Technology, under
contract with the National Aeronautics and Space Administration.

\clearpage

\clearpage

\begin{figure}
\plotone{Jordan.fig3.ps}
\figcaption[Jordan.fig3.ps]{\textit{(Left Panel)}
Surface brightness profiles for NGC~4839 in $B$ (crosses) and $I$ (filled circles). The
squares and circles are respective $B$- and $r$-band profiles from J{\o}rgensen, 
Franx \& Kj{\ae}rgaard (1992). Crosses denote the photographic profile of Schombert (1986).
The dot-dashed lines are best-fit de Vaucouleurs laws in the range $1.4 < r^{1/4} < 1.8$.
\textit{(Right Panels)} Ellipticity ($top$), position angle ($middle$) and $B_4$ ($bottom$) 
as a function of galactocentric distance. The arrow in the middle panel indicates the position 
angle of the major axis of Abell 1656 (Rhee, van Haarlen \& Katgert 1991; Plionis 1994).
\label{fig:n4839sb}
}
\end{figure}


\begin{figure}
\plotone{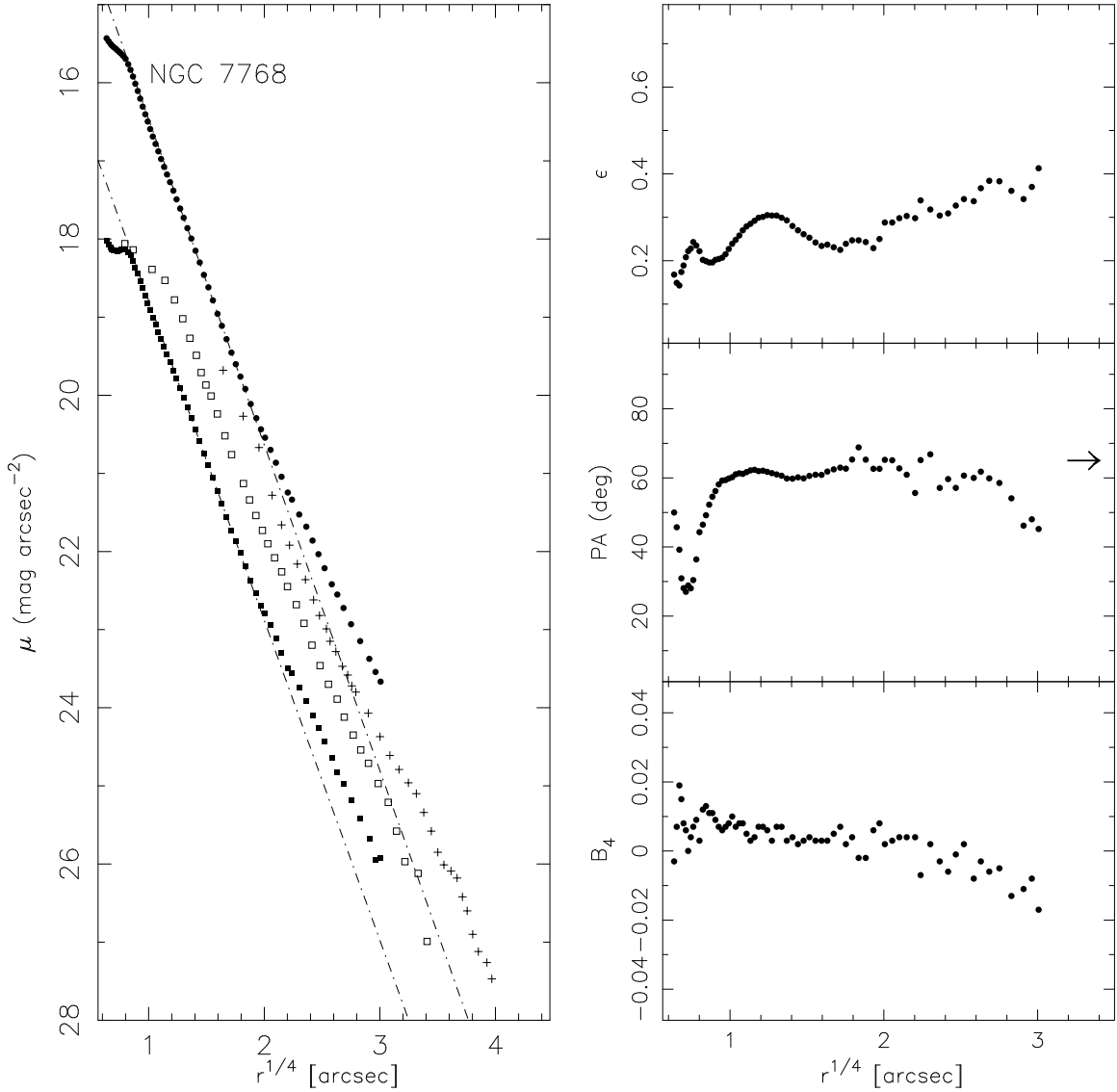}
\figcaption[Jordan.fig4.ps]{\textit{(Left Panel)}
Surface brightness profiles for NGC~7768 in $B$ (crosses) and $I$ (filled circles). 
The squares are a $V$-band profile from Malumuth \& Kirshner (1985), while the
crosses denote the photographic profile of Schombert (1986).
The dot-dashed lines are best-fit de Vaucouleurs laws in the range $1.4 < r^{1/4} < 1.8$.
Note the flattening inside $r \simeq 0\farcs5$ caused by the
central dust disk (see Figure~\ref{fig:nuc}).
\textit{(Right Panel)} Ellipticity ($top$), position angle ($middle$) and $B_4$ 
($bottom$) as a function of galactocentric distance. The arrow in the middle panel 
indicates the position angle of the major axis of Abell 2666 (Lambas, Groth \& 
Peebles 1988).
\label{fig:n7768sb}
}
\end{figure}

\clearpage

\begin{figure}
\plottwo{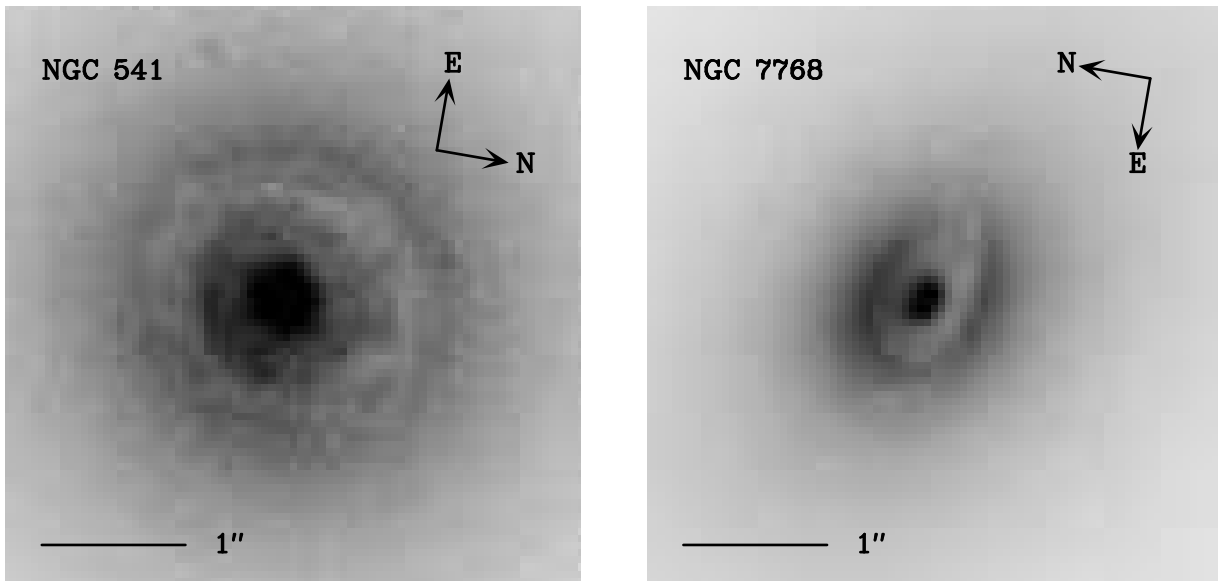}{Jordan.fig5b.ps}
\figcaption[nuc]{F450W images of the central regions of NGC~541 and NGC~7768. Note the 
obvious dust disks.
\label{fig:nuc}}
\end{figure}

\clearpage

\begin{figure}
\plotone{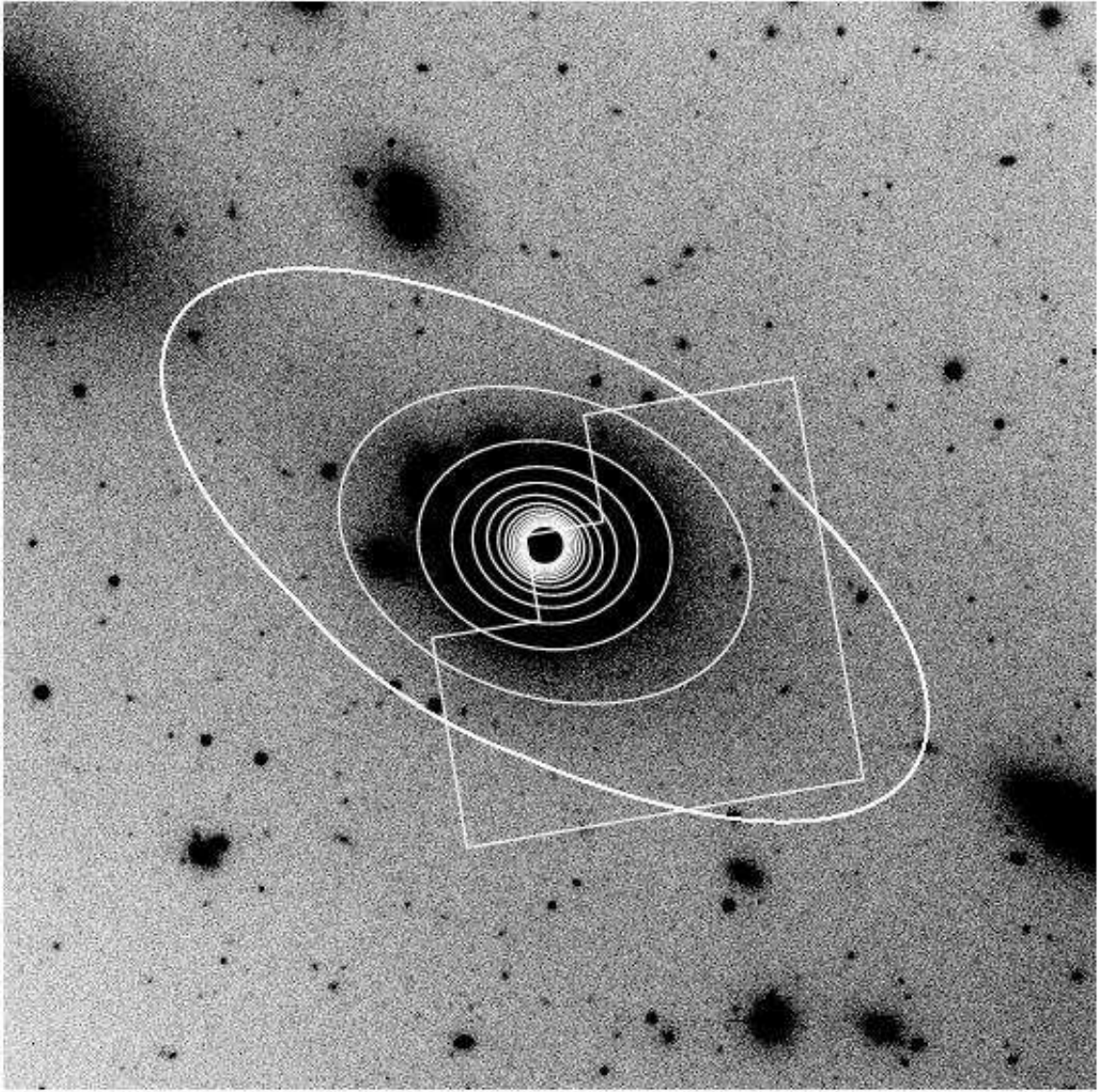}
\figcaption[galaxy.ps]{$I$-band image of NGC~541 taken with FORS1 at the VLT.
North is up and East is to the left in this image, which measures $6.8' \times 6.8'$.
Overlayed are the fitted ellipses for NGC~541, along with the WFPC2 field of view.
\label{fig:n541vlt}}
\end{figure}

\clearpage

\begin{figure}
\plotone{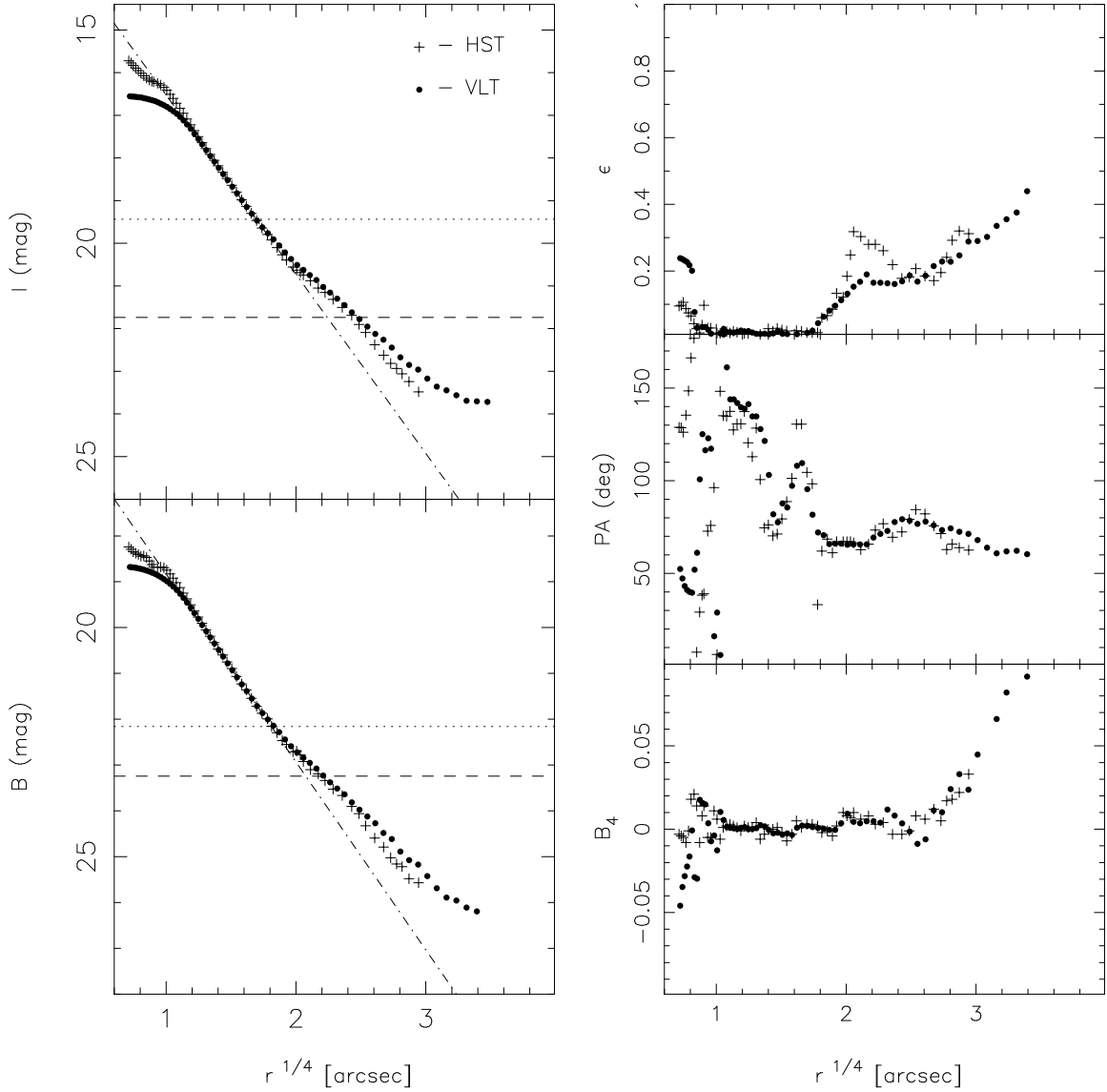}
\figcaption[Jordan.fig7.ps]{
\textit{(Left Panels)} Comparison of {\sl HST} and VLT surface photometry for 
NGC~541 in the $B$ (bottom) and $I$ (top) bandpasses. The dotted line
shows the sky brightness in the VLT frames; the dashed line is the
corresponding value in the {\sl HST} frames. The VLT profile was shifted to 
agree with that from {\sl HST} in the region $1.4 < r^{1/4} < 1.8$ ($i.e.,$ 
between 4$^{\prime\prime}$ and 10$^{\prime\prime}$). The dot-dashed
line is the best-fit de Vaucouleurs law in this range. The cD envelope is 
clearly visible as the excess above this line, beginning at 
$\approx 10^{\prime\prime}$.
\textit{(Right Panels)} Comparison of the ellipticity ($top$), position 
angle ($middle$) and $B_4$ ($bottom$) measurements as a function of $r^{1/4}$. 
\label{fig:vlthst}}
\end{figure}

\clearpage

\begin{figure}
\plotone{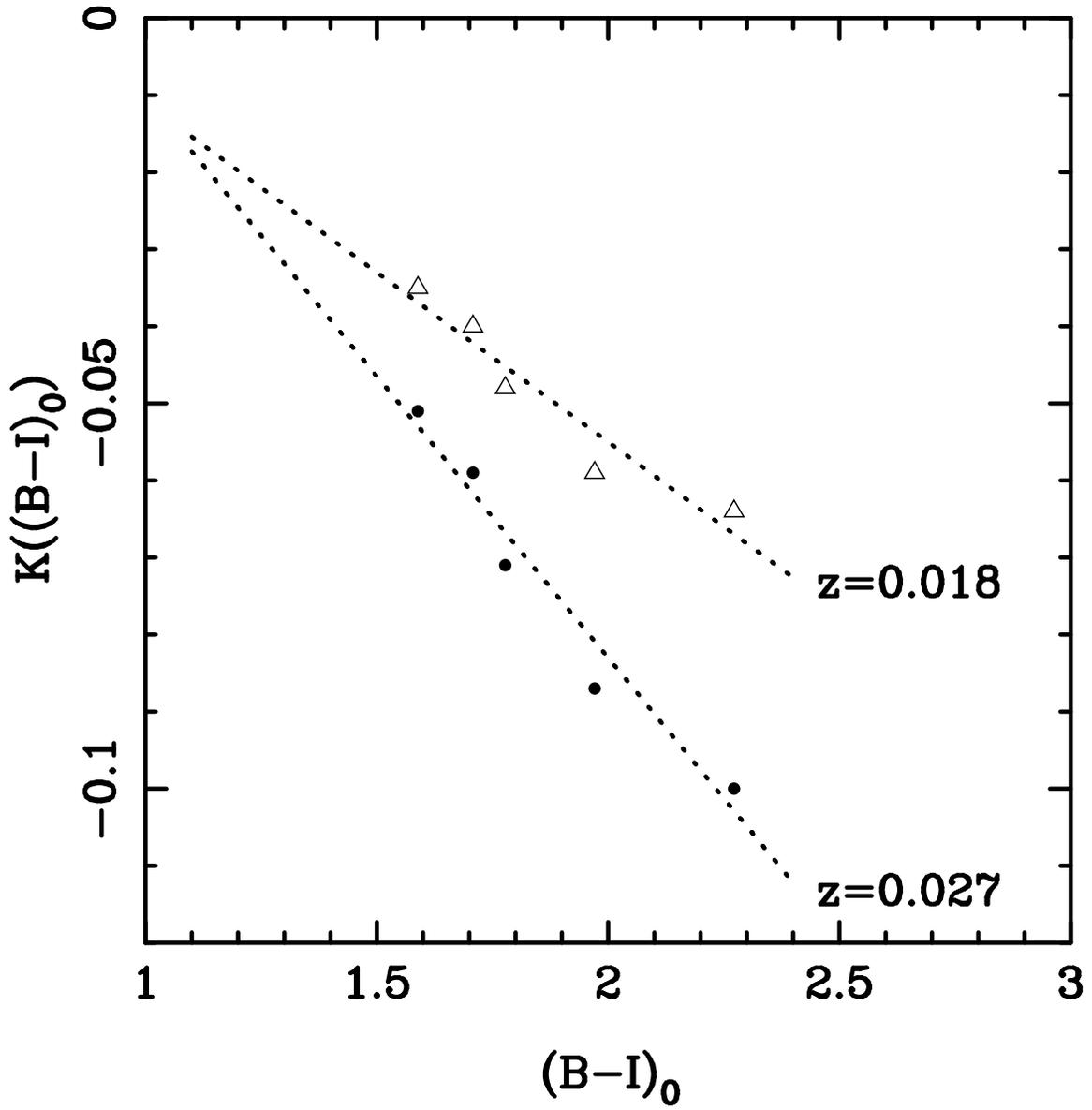}
\figcaption[Jordan.fig8.ps]{k-correction, $K(B-I)_0$, plotted as a function of 
dereddened rest frame color, $(B-I)_0$. Triangles and filled circles 
indicate the corrections for the galaxies with the lowest 
(NGC~541) and highest (NGC~7768) redshifts in our sample. The dotted 
lines are the adopted best-fit linear relations. Note the sizable 
extrapolation needed to correct the colors of the bluest objects.
\label{fig:kcorr}}
\end{figure}

\clearpage

\begin{figure}
\plotone{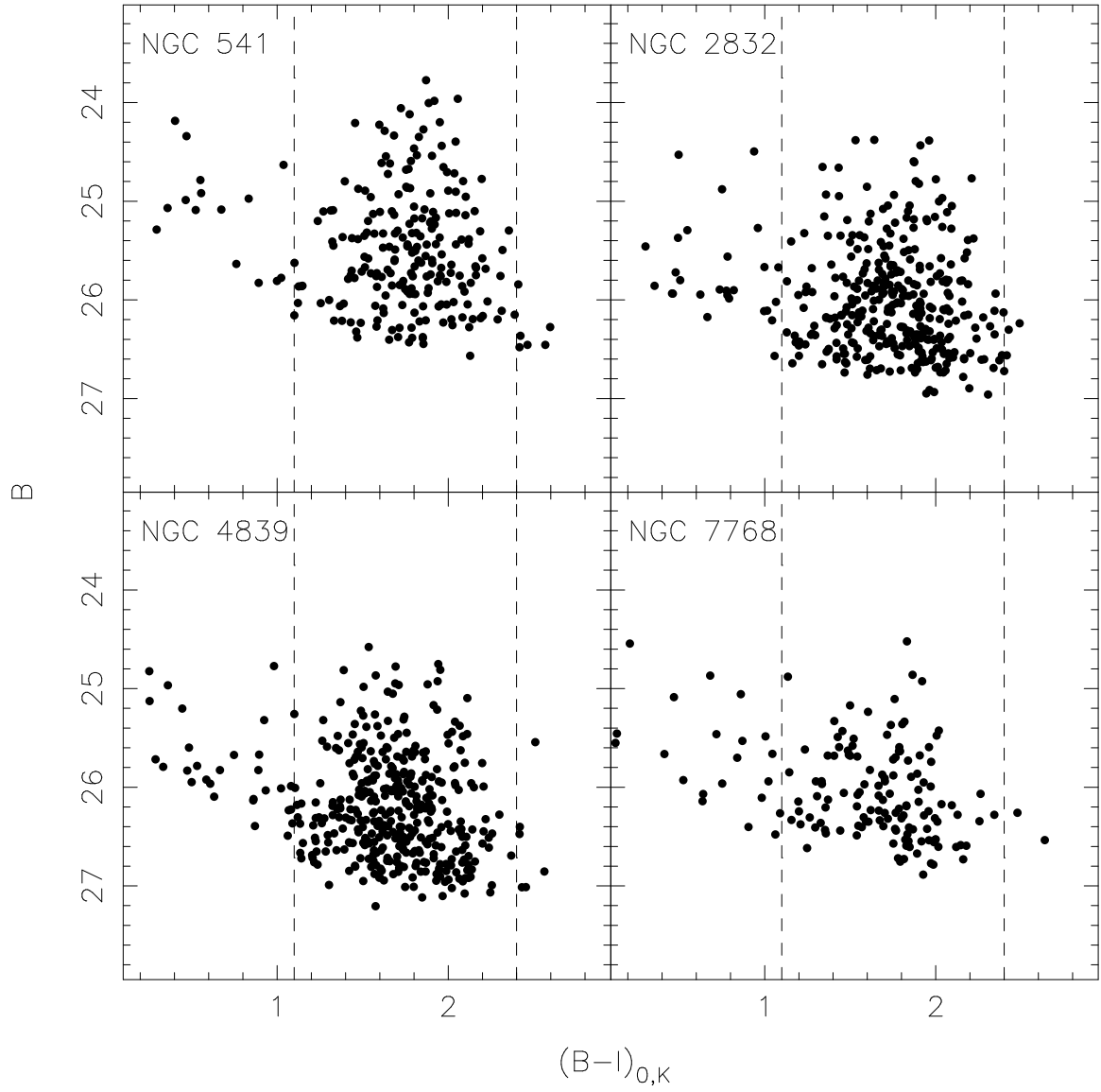}
\figcaption[Jordan.fig9.ps]{
Color magnitude diagrams for globular cluster candidates in our four program
galaxies. The magnitudes and colors have been corrected for reddening and 
extinction; k-corrections have been applied to only the latter. The dashed 
lines show the color range, $1.1 \le (B-I)_0 \le 2.4$, used to isolate
globular clusters from foreground stars and background galaxies.
\label{fig:cmd}}
\end{figure}

\clearpage

\begin{figure}
\plotone{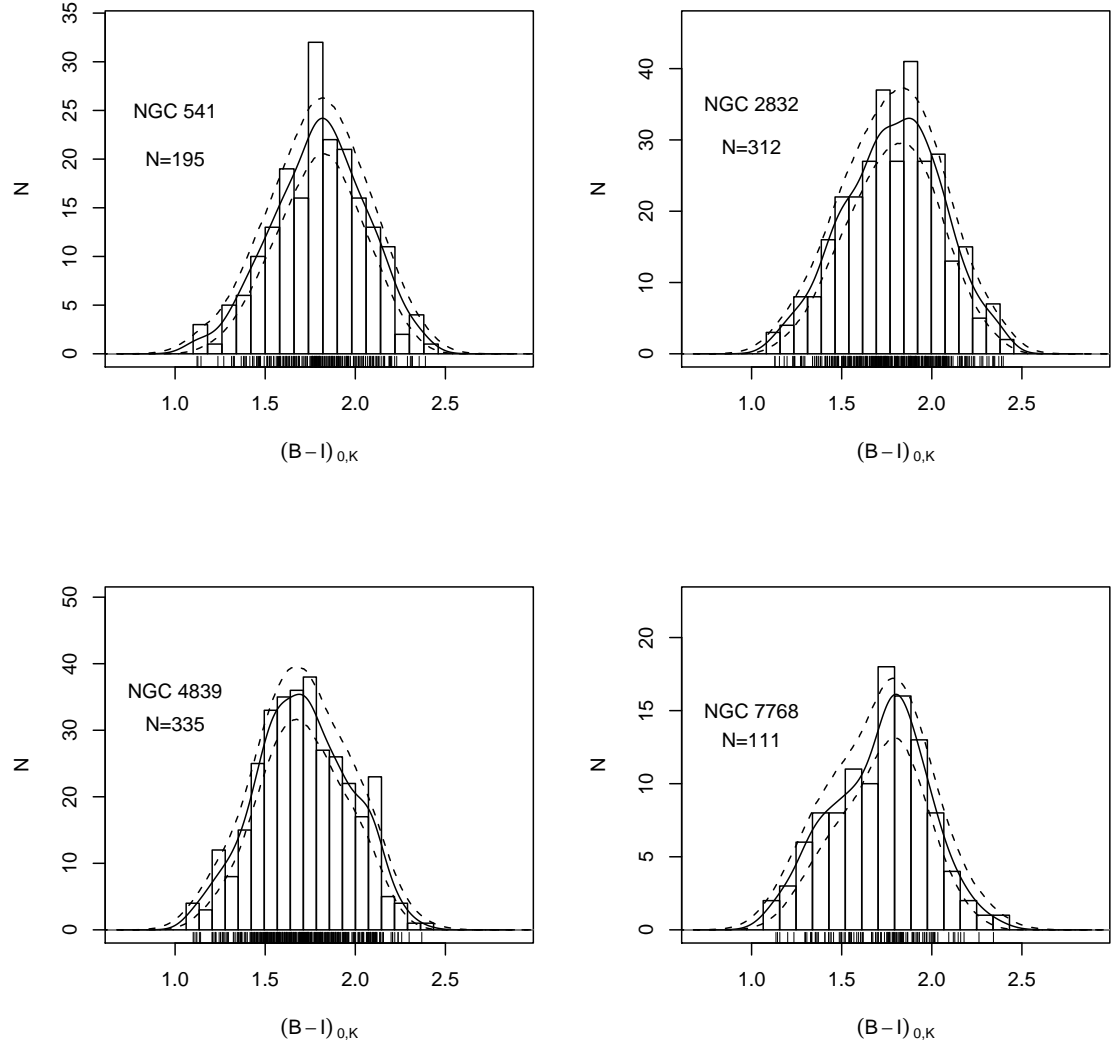}
\figcaption[Jordan.fig10.ps]{
Distribution of dereddened, k-corrected colors of globular cluster candidates
with color errors $\sigma{(B-I)} < 0.3$ mag. The number
of candidates is indicated in each panel. The solid curves are kernel density estimates 
of the color distributions, along with smoothed bootstrap estimates of their 90\% 
confidence bands (dashed curves). 
\label{fig:hists}}
\end{figure}

\clearpage

\begin{figure}
\plotone{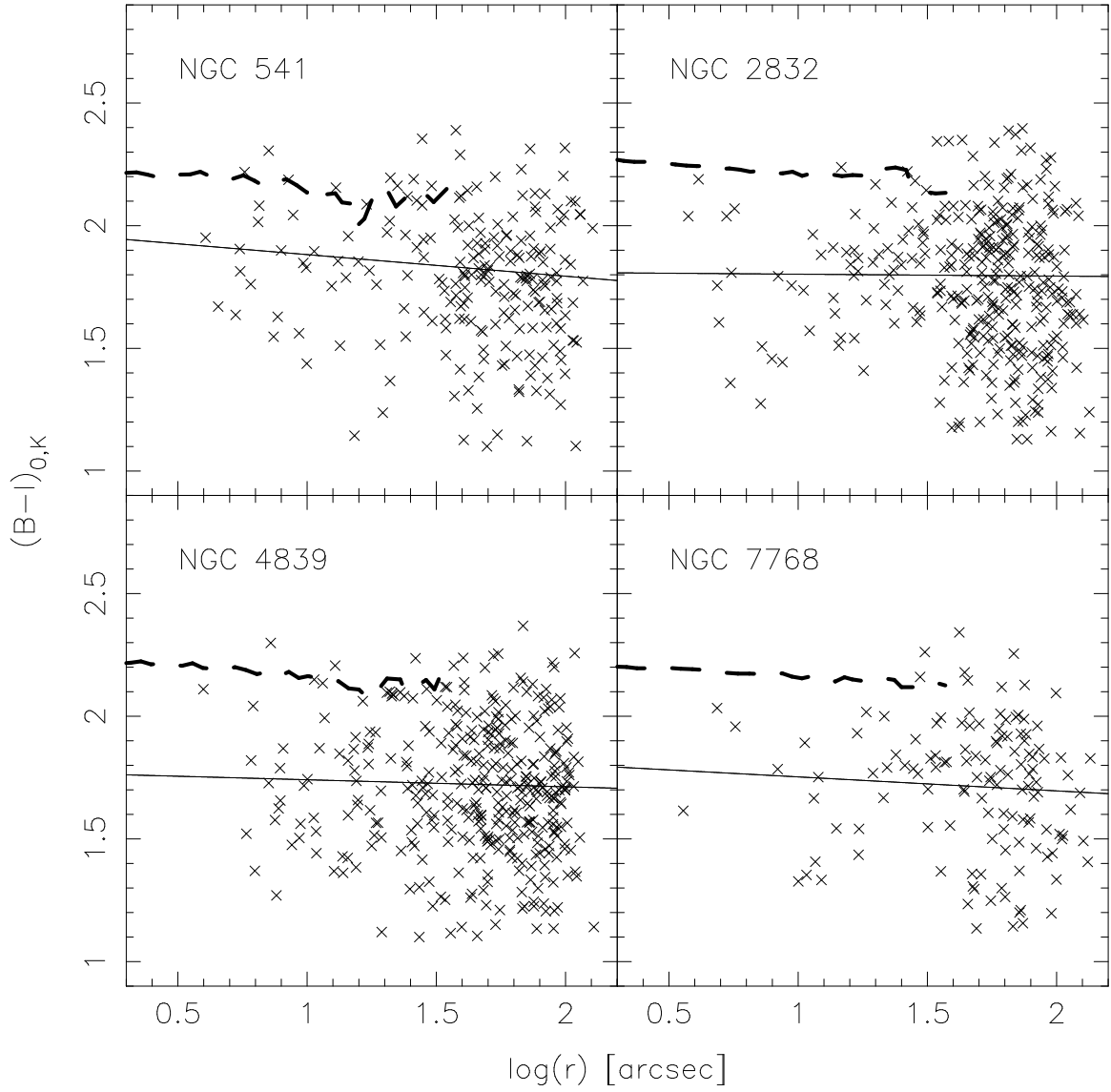}
\figcaption[Jordan.fig11.ps]{
Dereddened, $K$-corrected colors of globular cluster candidates plotted as a function of 
galactocentric radius for each of our program galaxies. The solid line is the line of 
best fit for objects with $\sigma{(B-I)} \leq 0.25$ mag. The dashed line is the color 
profile of the underlying galaxy light.
\label{fig:bivsr}}
\end{figure}

\clearpage

\begin{figure}
\plotone{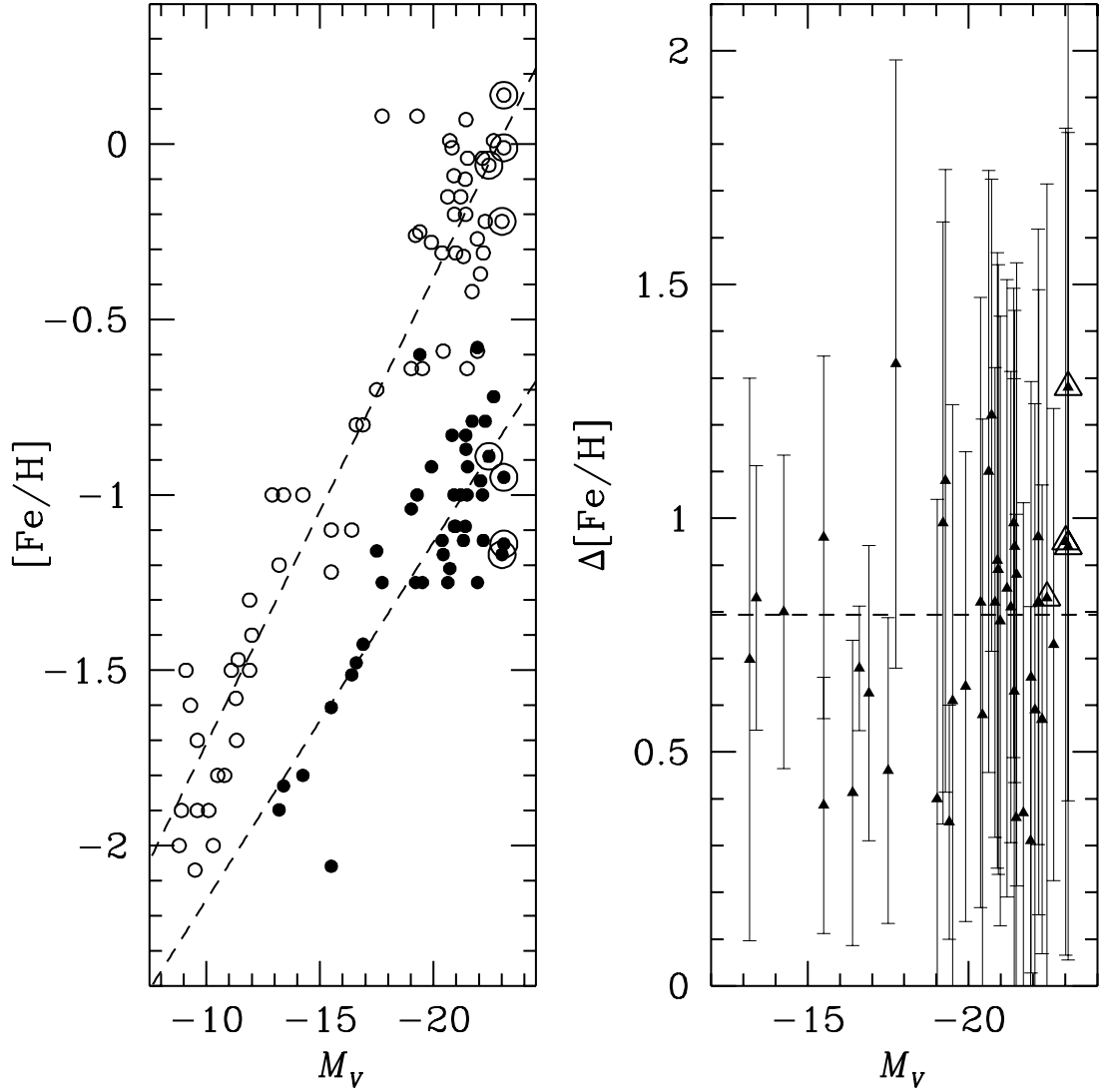}
\figcaption[Jordan.fig12.ps]{\textit{(Left Panel)} Mean metallicity of globular clusters (small filled 
circles) and stars (small open circles), plotted as a function of host galaxy absolute 
magnitude. Circled points indicate the four cD galaxies examined here.
The dashed lines are the lines of best-fit for the two samples.
\textit{(Right Panel)} Metallicity offset $\Delta$[Fe/H] between stars and globular 
clusters, plotted as a function of host galaxy absolute magnitude. The dashed line
indicates the weighted average: ${\Delta}$[Fe/H] = 0.79$\pm$0.04~dex.
Circled points indicate the four cD galaxies examined here.
\label{fig:metoffset}}
\end{figure}

\clearpage

\begin{figure}
\plotone{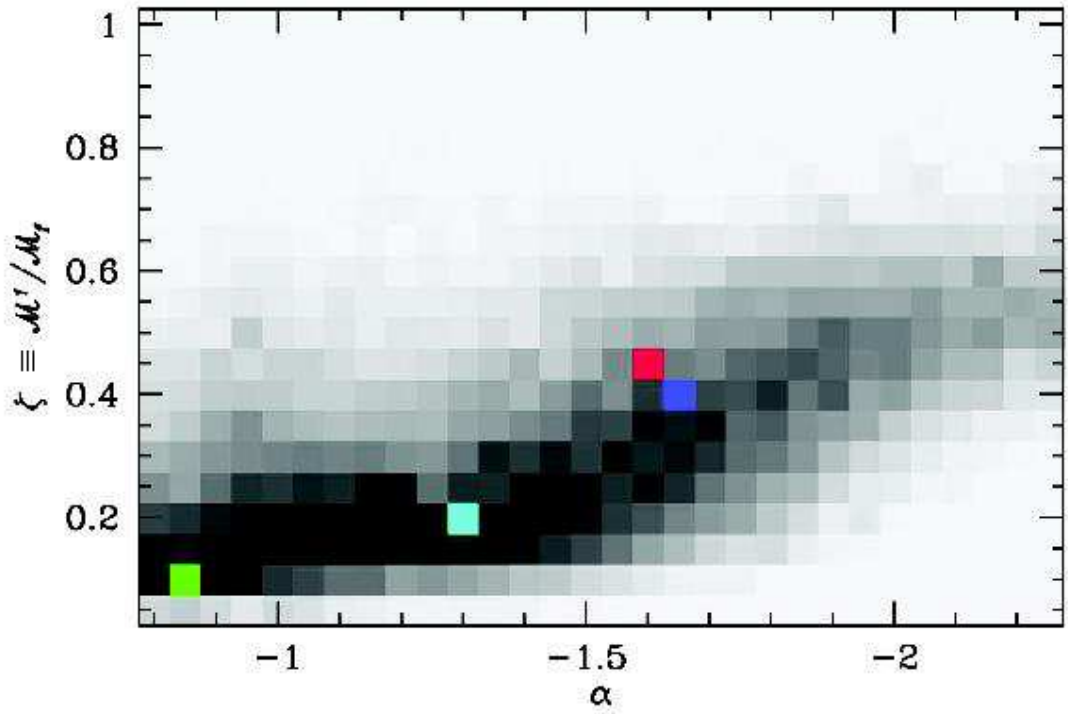}
\figcaption[pspec1.ps]{
Probability surface of ${\alpha}$-${\zeta}$ determined from our Monte Carlo simulations 
of the globular cluster metallicity distribution for NGC~541.  Each pixel in this
30$\times$20 image is proportional to the mean KS probability of 100 simulations for 
that ${\alpha}$-${\zeta}$ pair (in the sense that darker pixels have higher probabilities). 
The four coloured pixels mark the location in the ${\alpha}$-${\zeta}$ plane of the 
simulations shown in the next figure.
\label{fig:ks541}}
\end{figure}


\begin{figure}
\plotone{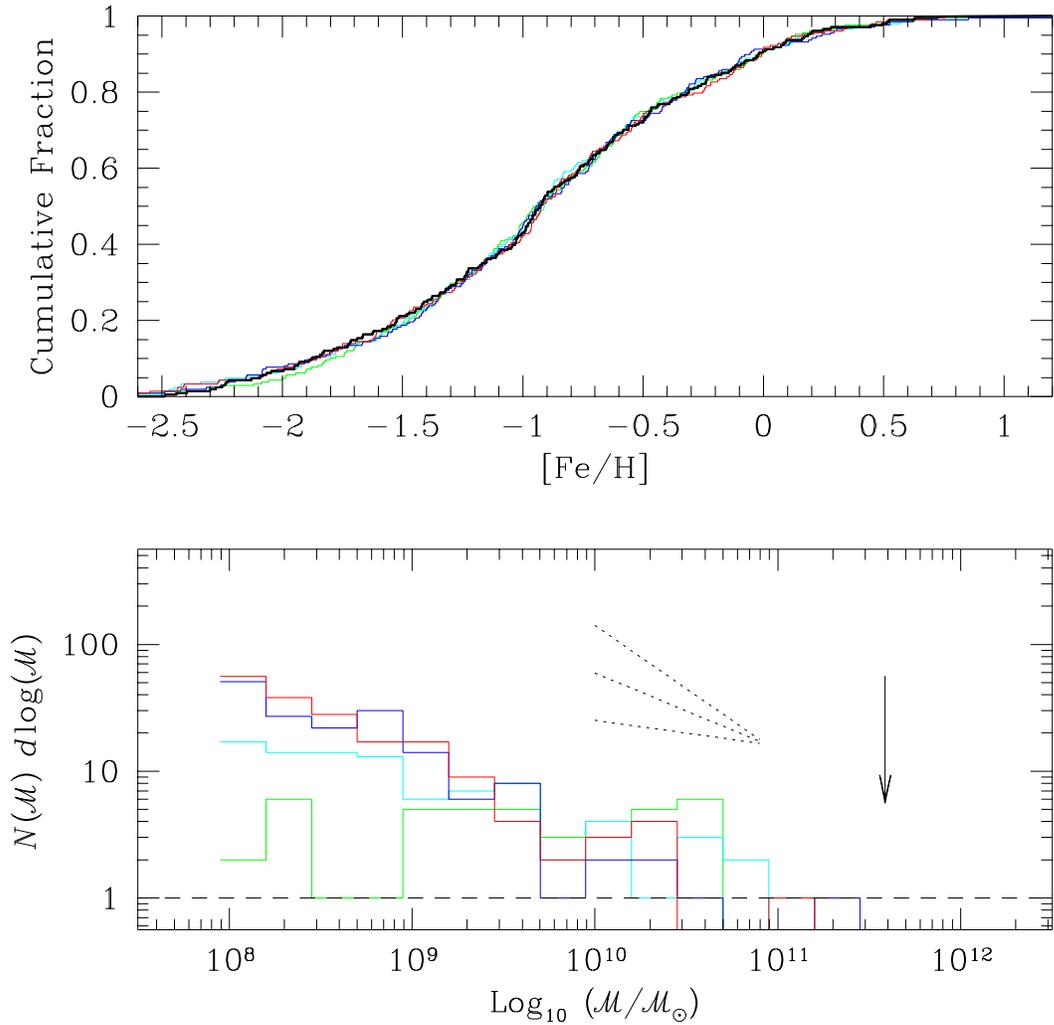}
\figcaption[Jordan.fig14.ps]{
{\it (Upper Panel)} Cumulative metallicity distribution of 208 globular cluster candidates 
associated with NGC541 (black curve). The four colored curves corresponds to the simulations
marked by the colored crosses in Figure~\ref{fig:ks541}.
In each case, a KS test indicates that there is a better than 99\%
probability that it was drawn from the same parent distribution as the real data.
{\it (Lower Panel)} Protogalactic mass spectra corresponding to the simulated metallicity 
distributions shown above.  The dotted lines show
powerlaw mass functions, $N({\cal M}) \propto {\cal M}^{-\gamma}$, with exponents of 
$\gamma = -2$, $-1.6$, and $-1.2$. The vertical arrow denotes the total mass of the 
final galaxy. A constant mass-to-light ratio of ${\Upsilon_V} = 5$ has been assumed 
in the conversion of luminosities to masses for both the final galaxy and the 
protogalactic fragments. 
\label{fig:mdf541}}
\end{figure}

\clearpage

\begin{figure}
\plotone{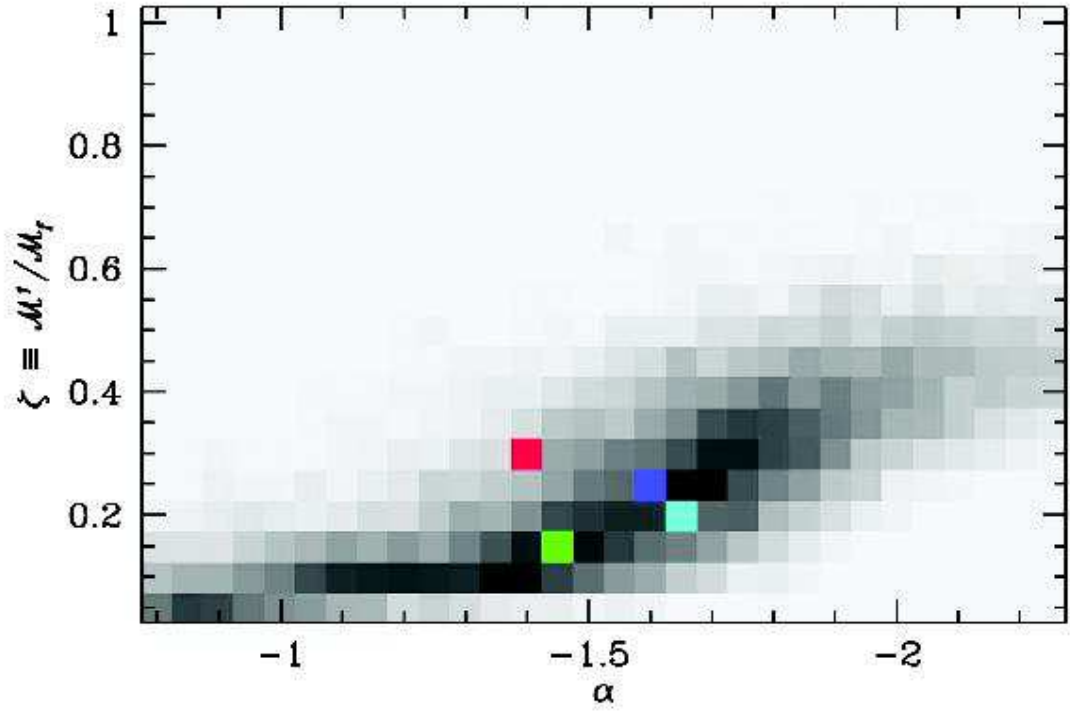}
\figcaption[pspec2.ps]{
Same as Figure~\ref{fig:ks541}, except for NGC~2832.
\label{fig:ks2832}}
\end{figure}


\begin{figure}
\plotone{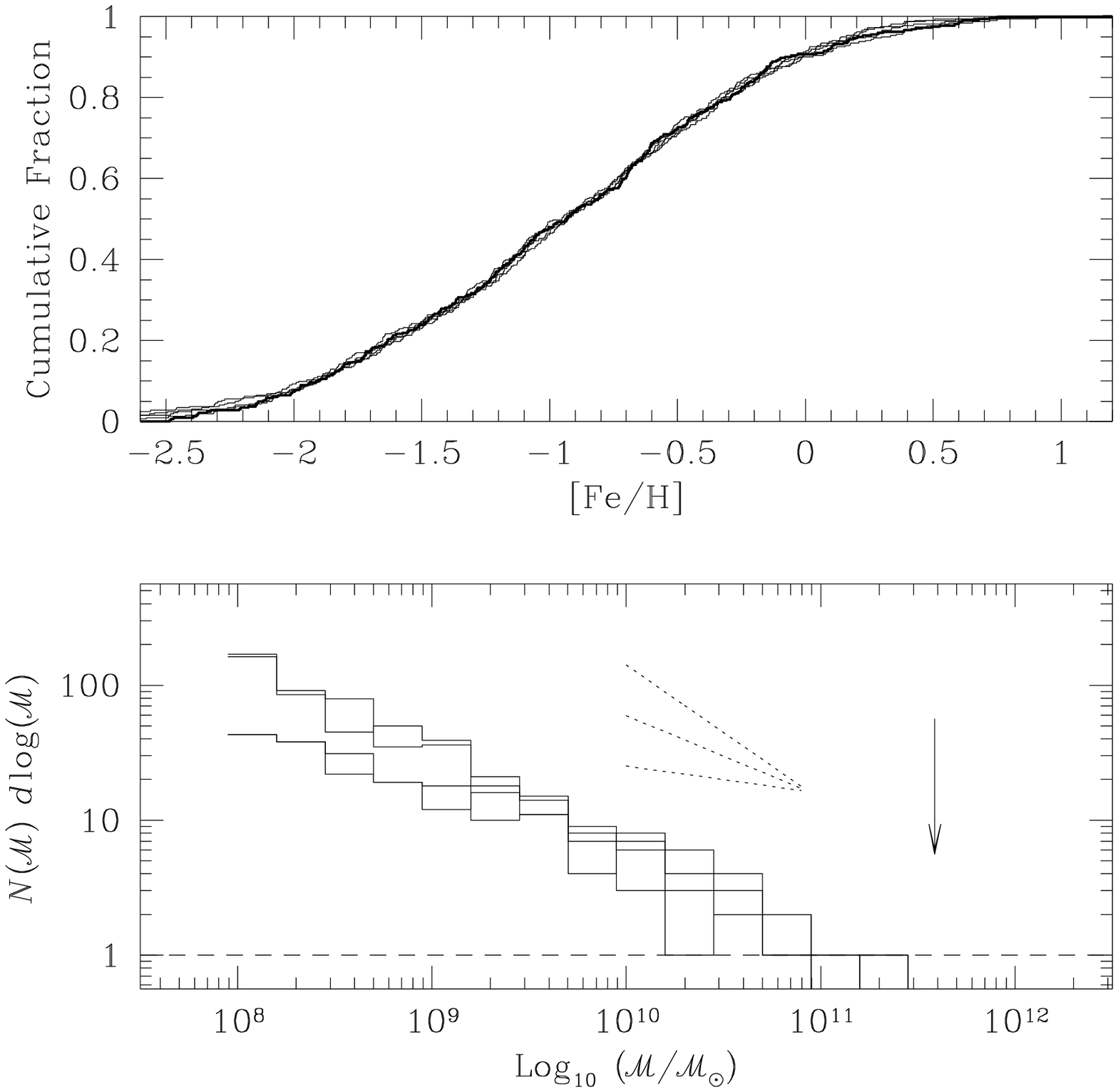}
\figcaption[Jordan.fig16.ps]{
Same as Figure~\ref{fig:mdf541}, except for NGC~2832. The cumulative metallicity distribution 
is based on a total of 323 candidate globular clusters.
\label{fig:mdf2832}}
\end{figure}

\clearpage

\begin{figure}
\plotone{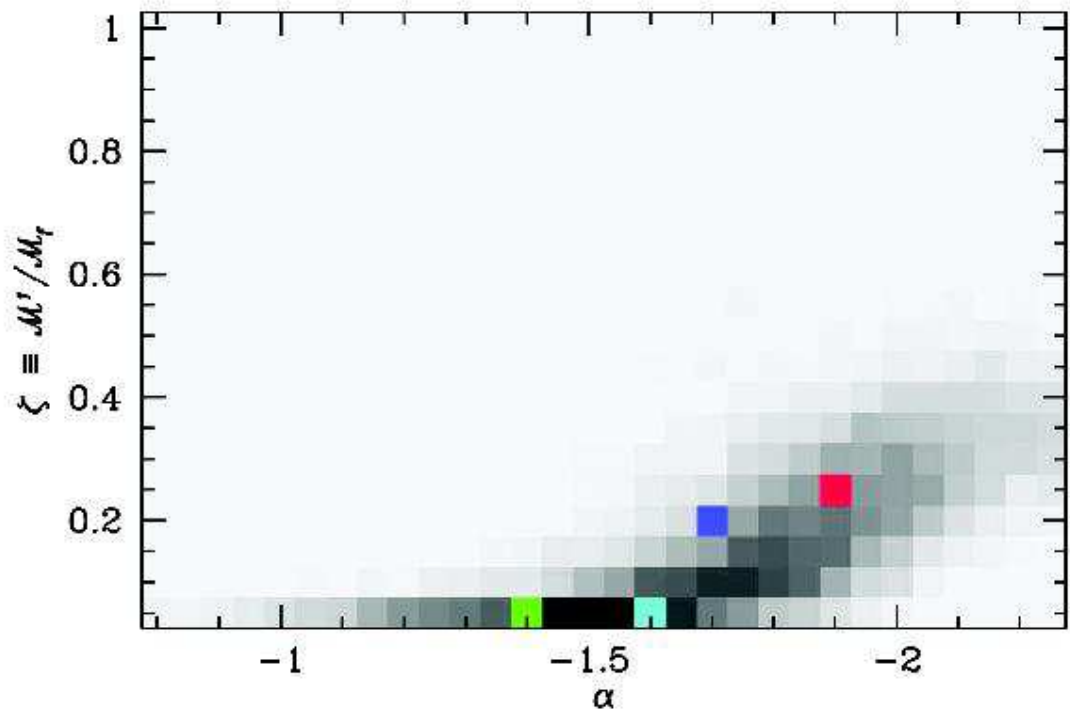}
\figcaption[pspec3.ps]{
Same as Figure~\ref{fig:ks541}, except for NGC~4839.
\label{fig:ks4839}}
\end{figure}


\begin{figure}
\plotone{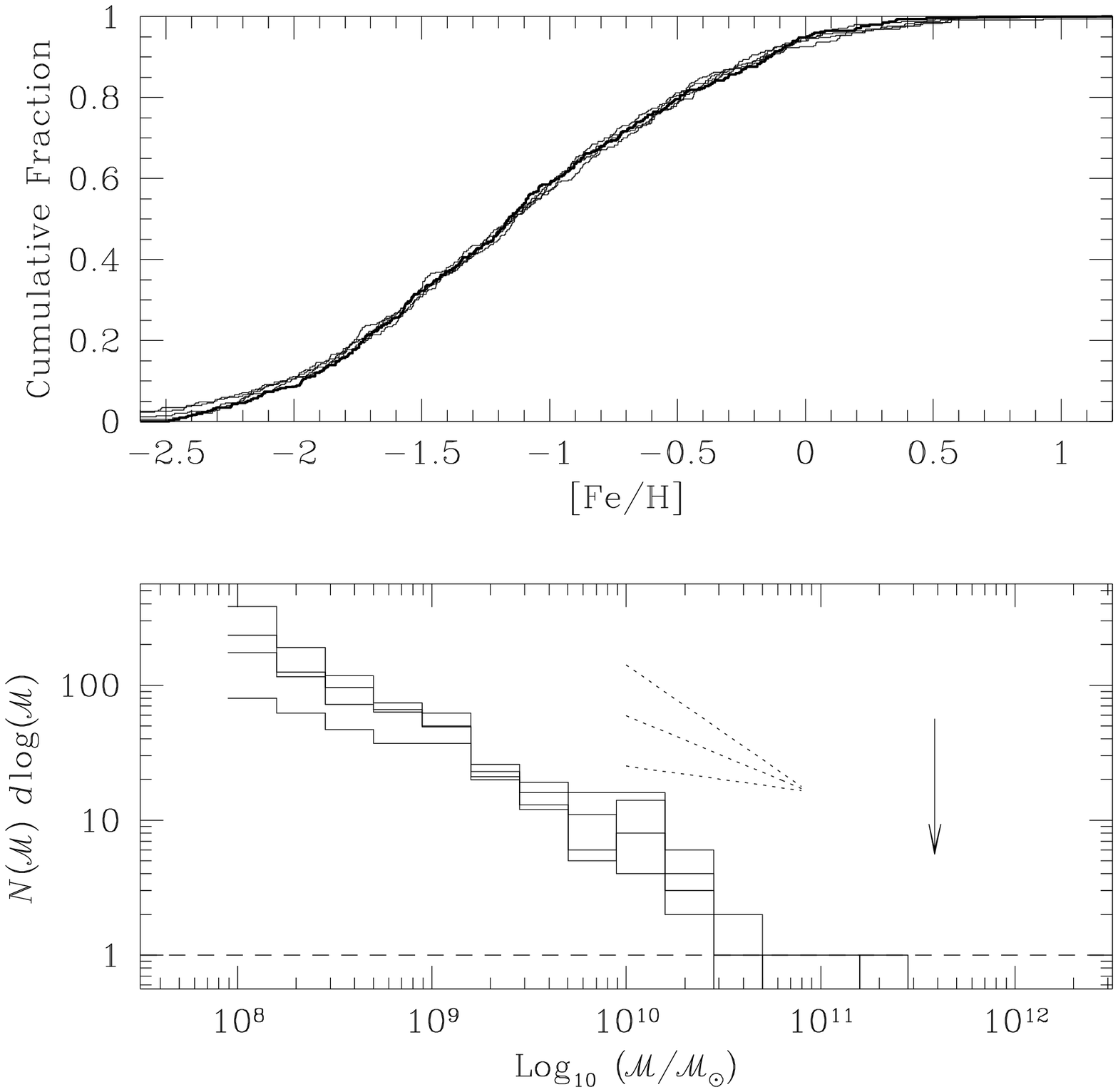}
\figcaption[Jordan.fig18.ps]{
Same as Figure~\ref{fig:mdf541}, except for NGC~4839. The cumulative metallicity distribution 
is based on total of 350 candidate globular clusters.
\label{fig:mdf4839}}
\end{figure}

\clearpage

\begin{figure}
\plotone{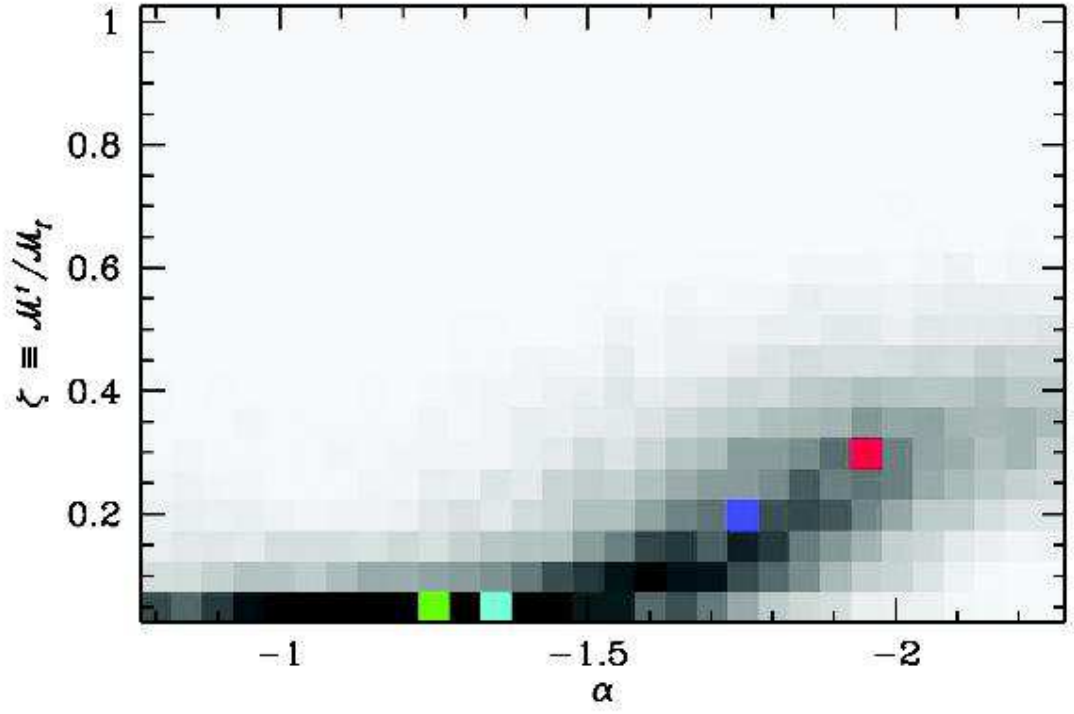}
\figcaption[pspec4.ps]{
Same as Figure~\ref{fig:ks541}, except for NGC~7768.
\label{fig:ks7768}}
\end{figure}


\begin{figure}
\plotone{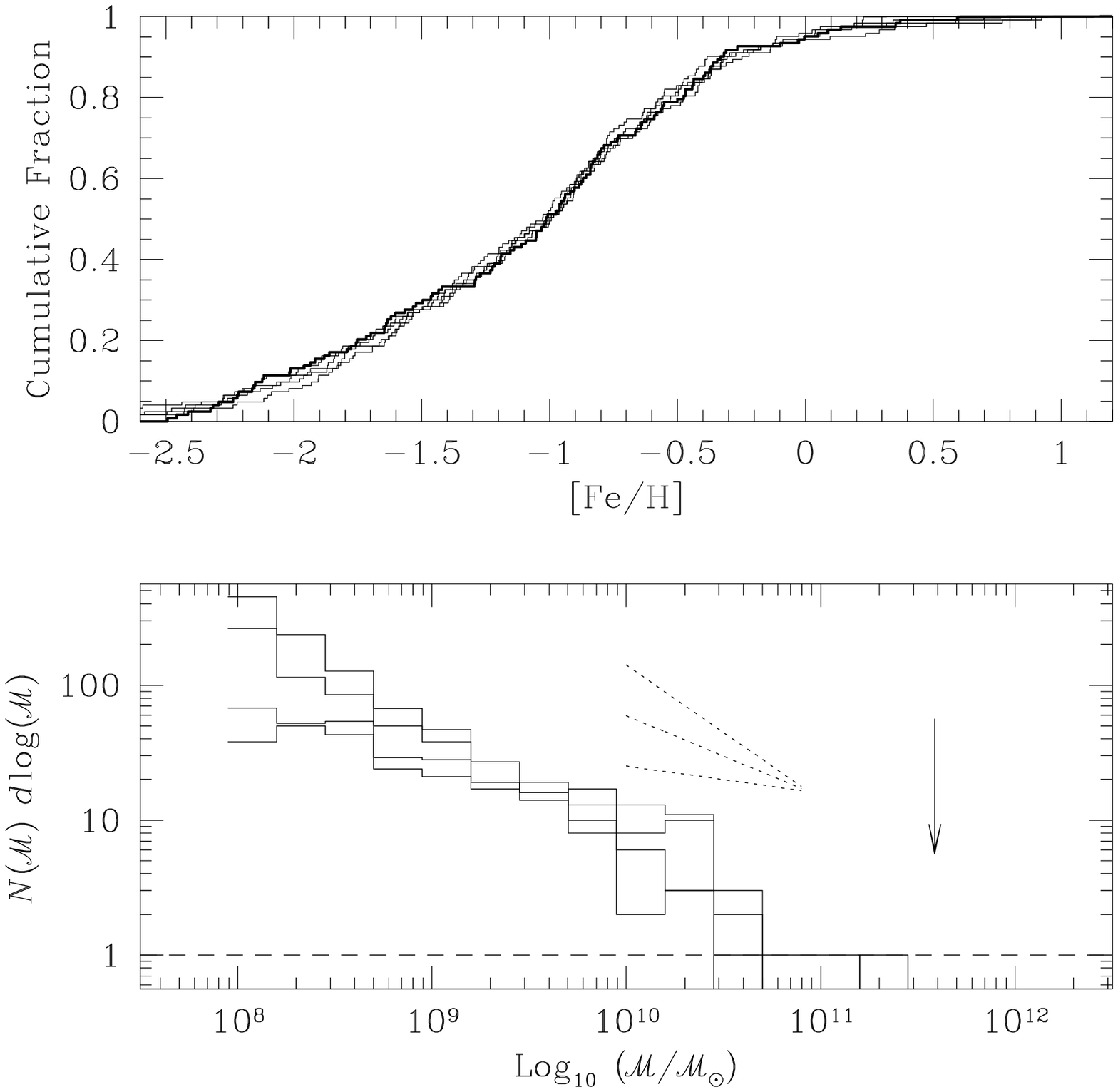}
\figcaption[Jordan.fig20.ps]{
Same as Figure~\ref{fig:mdf541}, except for NGC 7768. The cumulative metallicity distribution 
is based on total of 123 candidate globular clusters.
\label{fig:mdf7768}}
\end{figure}

\clearpage

\begin{deluxetable}{ccccc}
\tablecaption{Observing Log for GO 8184\label{tab:log}}
\tablewidth{0pt}
\tablehead{
\colhead{Galaxy} & \colhead{R.A.} & \colhead{Decl.}&
\colhead{Filter} & \colhead{Exposure Time}\\
\colhead{} & \colhead{(J2000)} & \colhead{(J2000)}&
\colhead{} & \colhead{(s)}
}
\startdata
NGC~~541  & 01:25:44.3 & --01:22:46 & F450W & 2$\times$1200\\
          &            &            &       & 1$\times$1000\\
          &            &            & F814W & 2$\times$~600\\
NGC~2832  & 09:19:46.9 &  +33:44:59 & F450W & 4$\times$1300\\
          &            &            & F814W & 2$\times$1300\\
NGC~4839  & 12:57:24.2 &  +27:29:54 & F450W & 6$\times$1300\\
          &            &            & F814W & 2$\times$1300\\
NGC~7768  & 23:50:58.6 &  +27:08:51 & F450W & 6$\times$1300\\
          &            &            &       & 1$\times$1200\\
          &            &            & F814W & 2$\times$1300\\
          &            &            &       & 1$\times$1200\\
\enddata
\end{deluxetable}


\begin{deluxetable}{lccccccc}
\scriptsize
\tablecolumns{8}
\tablewidth{0pt}
\tablecaption{Properties of Host Clusters\label{tab:prop_clus}}
\tablehead{
\colhead{Galaxy} &
\colhead{Cluster} &
\colhead{$\langle cz\rangle$\tablenotemark{a}} &
\colhead{$\sigma$\tablenotemark{a}} &
\colhead{$T_X$\tablenotemark{b}} &
\colhead{$L_X$\tablenotemark{b}} &
\colhead{Richness Class\tablenotemark{c}} &
\colhead{BM Class\tablenotemark{c}} \\
\colhead{} &
\colhead{} &
\colhead{(km s$^{-1}$)} &
\colhead{(km s$^{-1}$)} &
\colhead{(keV)} &
\colhead{(10$^{44}$ erg s$^{-1}$)} &
\colhead{} &
\colhead{} 
}
\startdata
N541   & A194  &5388$\pm$61  &  480$^{+48}_{-38}$   &2.63$\pm0.15$ & 0.22$\pm$0.04 &0       &II \\
N2832  & A779  &6887$\pm$110 &  503$^{+100}_{-63}$  &  \nodata     & 0.28$\pm$0.07 &0       &I-II \\
N4839  & A1656 &7339         &  329                 &4.5           &\nodata        &\nodata & \nodata \\
N7768  & A2666 &8086$\pm$105 &  476$^{+95}_{-60}$   &  \nodata     & 0.05          &0       &I \\
\enddata
\tablenotetext{a}{~$\langle cz\rangle $ and $\sigma$ from Zabludoff, Geller \& Huchra (1990),
except for NGC~4839 (Colless \& Dunn 1996).}
\tablenotetext{b}{~X-ray temperatures and luminosities for NGC~541 from Wu, Xue \& Fang (1999).
The X-ray temperature of NGC~4839 is taken from Neumann et~al. (2003).}
\tablenotetext{c}{~Richness and Bautz-Morgan classifications from Leir \& van den Bergh (1977).}
\end{deluxetable}


\begin{deluxetable}{rcccrr}
\tablecaption{Surface Brightness Profiles\label{tab:sb_all}}
\tablewidth{0pt}
\tablehead{
\colhead{$r$\tablenotemark{a}} & \colhead{$\mu_B$} & \colhead{$\mu_I$} & \colhead{$e$} &
\colhead{PA} & \colhead{$B_4$} \\
\colhead{(arcsec)} & \colhead{(mag arcsec$^{-2}$)} & \colhead{(mag arcsec$^{-2}$)} & \colhead{} & \colhead{(deg)} & \colhead{} 
}
\startdata
\multicolumn{6}{c}{NGC~~541} \nl
    0.16 &    15.507 &    17.998 &   0.07 &   120.6 &    -0.007 \\
    0.28 &    15.772 &    18.292 &   0.10 &   128.6 &    -0.004 \\
    0.42 &    16.012 &    18.442 &   0.07 &   166.3 &     0.018 \\
    0.58 &    16.175 &    18.546 &   0.01 &    29.1 &    -0.008 \\
    0.76 &    16.251 &    18.682 &   0.03 &    72.8 &    -0.005 \\
\enddata
\tablenotetext{a}{Semi-major axis distance}
\tablecomments{The complete version of this table is in the electronic
edition of the Journal.  The printed edition contains only a sample.}
\end{deluxetable}

\clearpage

\begin{deluxetable}{lccccc}
\tablecaption{GC candidates photometry for NGC~541 \label{tab:n541_gcphot}}
\tablewidth{0pt}
\tablehead{
\colhead{ID} & \colhead{$r$} &\colhead{$\Delta \alpha$} & \colhead{$\Delta \delta$} & \colhead{$B$} & \colhead{$(B-I)$} \\
\colhead{} & \colhead{(arcsec)} &\colhead{(arcsec)} & \colhead{(arcsec)} & \colhead{(mag)} & \colhead{(mag)} 
}
\startdata
   1 &     4.047 &    -4.012 &    -0.540 &    24.200 &     2.005\\
   2 &     4.523 &     0.087 &     4.522 &    25.102 &     1.711\\
   3 &     5.292 &    -5.029 &    -1.650 &    24.545 &     1.675\\
   4 &     5.489 &    -5.332 &    -1.308 &    25.594 &     1.956\\
   5 &     5.510 &    -5.369 &     1.243 &    25.321 &     1.861\\
\enddata
\tablecomments{The complete version of this table is in the electronic
edition of the Journal.  The printed edition contains only a sample.}
\end{deluxetable}


\begin{deluxetable}{lccccc}
\tablecaption{GC candidates photometry for NGC~2832 \label{tab:n2832_gcphot}}
\tablewidth{0pt}
\tablehead{
\colhead{ID} & \colhead{$r$} &\colhead{$\Delta \alpha$} & \colhead{$\Delta \delta$} & \colhead{$B$} & \colhead{$(B-I)$} \\
\colhead{} & \colhead{(arcsec)} &\colhead{(arcsec)} & \colhead{(arcsec)} & \colhead{(mag)} & \colhead{(mag)} 
}
\startdata
   1 &     3.756 &    -1.916 &     3.402 &    25.260 &     2.113\\
   2 &     4.117 &     2.033 &    -3.753 &    25.842 &     2.273\\
   3 &     4.865 &     4.644 &    -2.960 &    24.936 &     1.813\\
   4 &     4.941 &    -2.327 &    -4.546 &    25.486 &     1.653\\
   5 &     5.286 &     1.442 &     5.149 &    24.968 &     2.113\\
\enddata
\tablecomments{The complete version of this table is in the electronic
edition of the Journal.  The printed edition contains only a sample.}
\end{deluxetable}


\begin{deluxetable}{lccccc}
\tablecaption{GC candidates photometry for NGC~4839 \label{tab:n4839_gcphot}}
\tablewidth{0pt}
\tablehead{
\colhead{ID} & \colhead{$r$} &\colhead{$\Delta \alpha$} & \colhead{$\Delta \delta$} & \colhead{$B$} & \colhead{$(B-I)$} \\
\colhead{} & \colhead{(arcsec)} &\colhead{(arcsec)} & \colhead{(arcsec)} & \colhead{(mag)} & \colhead{(mag)} 
}
\startdata
   1 &     3.969 &     3.928 &     1.900 &    25.460 &     2.194\\
   2 &     5.818 &    -0.369 &    -5.809 &    25.387 &     1.565\\
   3 &     6.070 &     6.118 &     2.718 &    26.194 &     1.883\\
   4 &     6.182 &     5.539 &     3.752 &    25.980 &     2.121\\
   5 &     6.277 &    -6.548 &     2.381 &    25.137 &     1.402\\
\enddata
\tablecomments{The complete version of this table is in the electronic
edition of the Journal.  The printed edition contains only a sample.}
\end{deluxetable}


\begin{deluxetable}{lccccc}
\tablecaption{GC candidates photometry for NGC~7768 \label{tab:n7768_gcphot}}
\tablewidth{0pt}
\tablehead{
\colhead{ID} & \colhead{$r$} &\colhead{$\Delta \alpha$} & \colhead{$\Delta \delta$} & \colhead{$B$} & \colhead{$(B-I)$} \\
\colhead{} & \colhead{(arcsec)} &\colhead{(arcsec)} & \colhead{(arcsec)} & \colhead{(mag)} & \colhead{(mag)} 
}
\startdata
   1 &     3.593 &    -2.871 &     2.526 &    25.824 &     1.670\\
   2 &     4.855 &     1.898 &    -4.552 &    26.172 &     2.120\\
   3 &     5.717 &     5.231 &     3.319 &    26.242 &     2.038\\
   4 &     8.337 &     9.357 &    -0.428 &    25.635 &     1.853\\
   5 &    10.016 &     1.635 &     9.910 &    26.362 &     1.361\\
\enddata
\tablecomments{The complete version of this table is in the electronic
edition of the Journal.  The printed edition contains only a sample.}
\end{deluxetable}


\begin{deluxetable*}{cccccc}
\tablecaption{K-corrections\label{tab:kcorr}}
\tablewidth{0pt}
\tablehead{
\colhead{Template} & \colhead{[Fe/H]} & \colhead{$cz$} & \colhead{$z$} & \colhead{$K(B)$} & \colhead{$K(I)$}\\
\colhead{} & \colhead{(dex)} &  \colhead{(km~s$^{-1}$)} &  \colhead{} & \colhead{(mag)} & \colhead{(mag)}
}
\startdata
Elliptical & $\sim$ 0 & 5422$\pm$~6 & $0.01809\pm0.00002$ & $0.085$ & $0.020$ \\
           &          & 6948$\pm$~6 & $0.02318\pm0.00002$ & $0.111$ & $0.025$ \\
           &          & 7362$\pm$12 & $0.02456\pm0.00004$ & $0.118$ & $0.027$ \\
           &          & 8192$\pm$16 & $0.02732\pm0.00005$ & $0.130$ & $0.031$ \\
G2         & $-0.4$   & 5422$\pm$~6 & $0.01809\pm0.00002$ & $0.072$ & $0.013$ \\
           &          & 6948$\pm$~6 & $0.02318\pm0.00002$ & $0.092$ & $0.017$ \\
           &          & 7362$\pm$12 & $0.02456\pm0.00004$ & $0.097$ & $0.018$ \\
           &          & 8192$\pm$16 & $0.02732\pm0.00005$ & $0.108$ & $0.021$ \\
G3         & $-1.0$   & 5422$\pm$~6 & $0.01809\pm0.00002$ & $0.052$ & $0.006$ \\
           &          & 6948$\pm$~6 & $0.02318\pm0.00002$ & $0.067$ & $0.007$ \\
           &          & 7362$\pm$12 & $0.02456\pm0.00004$ & $0.071$ & $0.007$ \\
           &          & 8192$\pm$16 & $0.02732\pm0.00005$ & $0.078$ & $0.008$ \\
G4         & $-1.5$   & 5422$\pm$~6 & $0.01809\pm0.00002$ & $0.044$ & $0.004$ \\
           &          & 6948$\pm$~6 & $0.02318\pm0.00002$ & $0.056$ & $0.006$ \\
           &          & 7362$\pm$12 & $0.02456\pm0.00004$ & $0.059$ & $0.006$ \\
           &          & 8192$\pm$16 & $0.02732\pm0.00005$ & $0.066$ & $0.007$ \\
G5         & $-1.9$   & 5422$\pm$~6 & $0.01809\pm0.00002$ & $0.038$ & $0.003$ \\
           &          & 6948$\pm$~6 & $0.02318\pm0.00002$ & $0.048$ & $0.004$ \\
           &          & 7362$\pm$12 & $0.02456\pm0.00004$ & $0.050$ & $0.004$ \\
           &          & 8192$\pm$16 & $0.02732\pm0.00005$ & $0.057$ & $0.005$ \\
\enddata
\end{deluxetable*}


\begin{deluxetable}{lcc}
\tablecaption{K-correction Coefficients for $(B-I)_0$ \label{tab:lincons}}
\tablewidth{0pt}
\tablehead{
\colhead{Galaxy} & \colhead{$C_K$} & \colhead{$D_K$} \\
\colhead{} & \colhead{} & \colhead{(mag)} 
}
\startdata
NGC~~541 & $0.96$ & $0.032$ \\
NGC~2832 & $0.94$ & $0.049$ \\
NGC~4839 & $0.94$ & $0.056$ \\
NGC~7768 & $0.93$ & $0.059$ \\ 
\enddata
\end{deluxetable}


\begin{deluxetable*}{ccccc}
\tablecaption{Statistics of the Globular Cluster Color Distributions\label{tab:stats}}
\tablewidth{0pt}
\tablehead{
\colhead{Galaxy} & \colhead{$C_{BI}$\tablenotemark{a}}& \colhead{$S_{BI}$\tablenotemark{b}} & \colhead{$AI$\tablenotemark{c}} & \colhead{$TI$\tablenotemark{d}} \\
\colhead{} & \colhead{(mag)} & \colhead{(mag)} & \colhead{} & \colhead{}
}
\startdata
NGC~~541 & $1.81^{+0.02}_{-0.01}$ & $0.28^{+0.01}_{-0.01}$ & $0.24^{+0.49}_{-0.61}$ & $1.02^{+0.09}_{-0.05}$ \\
NGC~2832 & $1.79^{+0.02}_{-0.01}$ & $0.28^{+0.01}_{-0.01}$ & $0.15^{+0.41}_{-0.61}$ & $1.02^{+0.04}_{-0.08}$ \\
NGC~4839 & $1.71^{+0.02}_{-0.01}$ & $0.28^{+0.01}_{-0.01}$ & $0.21^{+0.74}_{-0.48}$ & $1.01^{+0.07}_{-0.04}$ \\
NGC~7768 & $1.72^{+0.02}_{-0.01}$ & $0.28^{+0.01}_{-0.02}$ & $0.47^{+0.28}_{-0.83}$ & $0.95^{+0.15}_{-0.03}$ \\
\enddata
\tablenotetext{a}{$C_{BI}$ = biweight location estimator (Beers, Flynn \& Gebhardt 1990).}
\tablenotetext{b}{$S_{BI}$ = biweight scale estimator (Beers, Flynn \& Gebhardt 1990).}
\tablenotetext{c}{$AI$ = assymetry index (Finch 1977).}
\tablenotetext{d}{$TI$ = tail index (Hoaglin, Mosteller \& Tukey 1983).}
\end{deluxetable*}


\begin{deluxetable*}{lcccccccc}
\scriptsize
\tablecolumns{8}
\tablewidth{0pt}
\tablecaption{Observed and Derived Photometric Properties of cD Galaxies\label{tab:colorinfo}}
\tablehead{
\colhead{Galaxy} &
\colhead{$E(B-I)$} &
\colhead{$\langle B-I \rangle$$^{\rm gal}_{\rm 0,k}$} &
\colhead{$\langle B-I \rangle$$^{\rm GC}_{\rm 0,k}$} &
\colhead{$V_{\rm tot}$} &
\colhead{$cz$} &
\colhead{$M_{\rm V,tot}$} &
\colhead{$L$/$L^{*}$\tablenotemark{a}}\\
\colhead{} &
\colhead{(mag)} &
\colhead{(mag)} &
\colhead{(mag)} &
\colhead{(mag)} &
\colhead{(km s$^{-1}$)} &
\colhead{(mag)} &
\colhead{} &
}
\startdata
NGC~~541 & 0.080 & 2.14 & 1.81 & 12.08$\pm$0.15 & 5422$\pm$~6 & $-$22.44$\pm$0.28 & 3.9 \\
NGC~2832 & 0.039 & 2.15 & 1.79 & 11.87$\pm$0.13 & 6948$\pm$~6 & $-$23.10$\pm$0.26 & 7.2 \\
NGC~4839 & 0.023 & 2.17 & 1.71 & 12.06$\pm$0.10 & 7362$\pm$12 & $-$23.02$\pm$0.25 & 6.7 \\
NGC~7768 & 0.087 & 2.22 & 1.72 & 12.30$\pm$0.13 & 8191$\pm$16 & $-$23.10$\pm$0.26 & 7.2 \\
\enddata
\tablenotetext{a}{For $M_R = -21.79 + 5\log{h_{65}}$ (Mobasher et~al. 2003), 
$H_0 = 72$~km~s$^{-1}$~Mpc$^{-1}$ (Freedman et~al. (2001), and $\langle V-R \rangle = 0.61$
Fukugita, Shimasaku \& Ichikawa (1995).}
\end{deluxetable*}


\begin{deluxetable}{lccc}
\scriptsize
\tablecolumns{4}
\tablewidth{0pt}
\tablecaption{Globular Cluster Specific Frequencies\label{tab:snval}}
\tablehead{
\colhead{Galaxy} &
\colhead{$M_{\rm V}^{\rm met}$}  &
\colhead{$N_{\rm GC}^{\rm met}$} &
\colhead{$S_{\rm N}^{\rm met}$}  \\
\colhead{} &
\colhead{(mag)} &
\colhead{} &
\colhead{}
}
\startdata
NGC~~541 & $-21.40\pm0.2$ & 1620$\pm$450 & $4.4\pm1.5$\\
NGC~2832 & $-22.35\pm0.2$ & 3200$\pm$920 & $3.7\pm1.3$\\
NGC~4839 & $-21.84\pm0.2$ & 3060$\pm$850 & $5.6\pm1.9$\\
NGC~7768 & $-22.33\pm0.2$ & 1850$\pm$570 & $2.2\pm0.8$\\
\enddata
\end{deluxetable}

\end{document}